\documentclass[]{article}
\usepackage[a4paper, total={6in, 10in}]{geometry}
\usepackage{authblk}
\usepackage[pagebackref=true]{hyperref} %for hyperlinks to work. pagebackref activates links back to citations in text from references.
\usepackage{amsmath} %align works only if this is there
\usepackage{amssymb} %this is so I can get 'therefore' symbol and leftrightarrow
\usepackage{cancel} %this is for cancelling terms
\usepackage{xcolor} %for colourful text
\usepackage{mathtools} %for colon equals symbol
\usepackage{dsfont} % for double stroked number font used at one place for 1
\usepackage{xparse} %for more than one optional argument in newcommand
\usepackage{cite}
\hypersetup{
	unicode=false,          
	pdftoolbar=true,        
	pdfmenubar=true,        
	pdffitwindow=true,     % window fit to page when opened
	pdfstartview={FitH},    % fits the width of the page to the window
	pdftitle={Spin 1 ERG-Hol RG},    % title
	pdfauthor={Pavan Dharanipragad},     % author
	pdfsubject={AdS/CFT},   % subject of the document
	pdfcreator={Pavan Dharanipragada},   % creator of the document
	pdfproducer={Pavan Dharanipragad}, % producer of the document
	pdfkeywords={Holographic RG} {ERG} {AdS/CFT}, % list of keywords
	pdfnewwindow=true,      % links in new window
	colorlinks=true,        % false: boxed links; true: colored links
	linkcolor=red,          % color of internal links
	citecolor=blue,         % color of links to bibliography
	filecolor=magenta,      % color of file links
	urlcolor=blue,         % color of external links
	linktocpage=true
}
\newcommand{\Dt}{\frac{D}{2}}
\newcommand{\lm}{\Lambda}
\newcommand{\hf}{\frac{1}{2}}
\newcommand{\eps}{\epsilon}
\newcommand{\CD}{{\cal D}}
\newcommand{\ddz}[1][]{\frac{\partial #1}{\partial z}}

\newcommand{\pmu}{\frac{\partial}{\partial x^\mu}}
\newcommand{\pMU}{\partial^{\mu}}
\newcommand{\pnu}{\frac{\partial}{\partial y^\nu}}

\newcommand{\Dmu}{D_{\mu}}
\newcommand{\DMU}{D^{\mu}}
\newcommand{\Amu}{A_{\mu}}
\newcommand{\AMU}[1][]{A^{\mu #1}}
\newcommand{\fd}[1]{\frac{\delta \hfill}{\delta #1}}
\newcommand{\cmd}[1][i]{\chi_{\mu}^{#1}}
\newcommand{\cmu}[1][i]{\chi^{\mu}_{ #1}}
\newcommand{\cnd}[1][i]{\chi_{\nu}^{#1}}
\newcommand{\cnu}[1][i]{\chi^{\nu }_{#1}}

\newcommand{\smu}[1][i]{\sigma^{\mu}_{#1}}
\newcommand{\smd}[1][i]{\sigma_\mu^{#1}}
\newcommand{\snu}[1][i]{\sigma^{\nu}_{#1}}
\newcommand{\snd}[1][i]{\sigma_\nu^{#1}}
\newcommand{\sru}[1][k]{\sigma^{\rho}_{#1}}
\newcommand{\srd}[1][k]{\sigma_\rho^{#1}}
\newcommand{\dsmu}[1][i]{\dot{\sigma}^{\mu}_{#1}}
\newcommand{\dsmd}[1][i]{\dot{\sigma}_{\mu}^{#1}}
\newcommand{\sqn}{\frac1{\sqrt N}}
\newcommand{\fih}[1][Ia]{\phi_h^{#1}}
\newcommand{\fil}[1][Ia]{\phi_l^{#1}}
\newcommand{\dhi}{\Delta_{h}^{-1}}

\newcommand{\Dl}[1][xy]{\Delta_{l#1}}
\newcommand{\dli}{\Delta_l^{-1}}
\newcommand{\dij}[1][ij]{\delta_{#1}}

\newcommand{\K}[1][ab]{K_{#1}}
\newcommand{\dxy}{\delta(x-y)}
\newcommand{\sil}{S_{I,\Lambda}}
\DeclareMathOperator{\Tr}{Tr}
\newcommand{\ddt}{\frac{\partial\hfill}{\partial t}}
\newcommand{\ddvar}[2]{\frac{\partial}{\partial #1^{#2}}}

\newcommand{\e}{\mathrm e^}
\newcommand{\I}{I_{\mu\nu}}
\newcommand{\dmn}{\delta_{\mu\nu}}
\newcommand{\dumn}{\delta^{\mu\nu}}
\newcommand{\pz}{\partial_z}
\newcommand{\amup}[1][i]{a^{\mu }_{#1}}
\newcommand{\amd}[1][i]{a_{\mu}^{#1}}

\newcommand{\der}{\mathrm d}
\newcommand{\sfu}{\sigma_{fi}^{\mu}}
\newcommand{\sbu}{\sigma_{bi}^{\mu}}
\newcommand{\sfd}{\sigma_{f\mu}^{ij}}
\newcommand{\dk}{\frac{\mathrm d^Dk}{(2\pi)^D}}
\newcommand{\skewsym}{Q^{IJ}}
\newcommand{\levciv}[1][ab]{\epsilon_{#1}}
\newcommand{\Levciv}[1][cd]{\epsilon^{#1}}
\newcommand{\field}{\phi^{Ia}}
\newcommand{\Qeps}{Q^{IJ}\epsilon_{ab}}
\NewDocumentCommand{\gen}{O{i} O{a} O{c}}{(T_{#1})^{#2}_{#3}}
\newcommand{\genca}{(T_i)^c_a}
\NewDocumentCommand{\dab}{O{a} O{c}}{\delta^{#1}_{#2}}
\newcommand{\eijk}{\epsilon^{ijk}}
\newcommand{\dpqr}{\delta^D(p+q+r)}
\newcommand{\sqG}{z^{-D-1}}

\title{Yang-Mills interaction from boundary vector model}
\author[1]{Pavan Dharanipragada \thanks{\href{mailto:pavan@physics.iitm.ac.in}{pavan@physics.iitm.ac.in}}}
\author[2,3,4]{B. Sathiapalan \thanks{\href{mailto:bala@imsc.res.in}{bala@imsc.res.in}}}

\affil[1]{Centre for Strings, Gravitation and Cosmology, Department of Physics, Indian Institute of Technology Madras,  Chennai 600036, India}
\affil[2]{The Institute of Mathematical Sciences, CIT Campus, Tharamani, Chennai 600113, India}
\affil[3]{Homi Bhabha National Institute, Training School Complex, Anushakti Nagar, Mumbai 400085, India}
\affil[4]{Chennai Mathematical Institute, H1, SIPCOT IT Park, Siruseri
	Kelambakkam 603103,
	India}

\begin{document}

\maketitle

{\centering \large \textit{Dedicated to A. P. Balachandran}\par}
\begin{abstract}
We construct a non-abelian spin 1 gauge theory with a cubic interaction in AdS$_4$ from the Exact Renormalisation Group (ERG) flow of a CFT in 3 dimensions. The latter is the ${USp}(2N)$ singlet sector of the free field theory of $2N$ massless complex scalars. The quadratic and cubic terms in the bulk action are those of a gauge fixed version of a (local) gauge invariant action. By construction this bulk action is the evolution operator for the ERG equation of the boundary theory and thus is guaranteed to reproduce the correct boundary correlators using the usual AdS/CFT prescription. This work expands on the programme first set out in \cite{sathiapalan_holographic_2017}. 
\end{abstract}
\tableofcontents

\section{Introduction}
The idea of holography is one of the most interesting ideas  in theoretical physics and it suggests that quantum gravity can be understood in terms of a lower dimensional non-gravitational field theory\cite{tHooft:1993dmi,Susskind:1994vu}. A very concrete realization of this idea is encoded in the AdS/CFT conjecture, for which there is ample evidence, according to which gravitational theories in AdS space are dual to CFT's in the boundary of AdS \cite{Maldacena:1997re,Gubser:1998bc,Witten:1998qj,Witten:1998zw}.
The radial direction of the bulk in AdS/CFT corresponds to the scale of the boundary theory. Thus the bulk radial evolution performs RG on the boundary theory.\footnote{The boundary theory is a CFT, but the boundary conditions can turn on deformations of the CFT action and then the boundary action will flow.} This idea is referred to as holographic renormalisation group \cite{Akhmedov:1998vf,Akhmedov:2002gq,Akhmedov:2010mz,Alvarez:1998wr,Balasubramanian:1999jd,Freedman:1999gp,deBoer:1999tgo,deBoer:2000cz,Faulkner:2010jy,Heemskerk:2010hk,Lizana:2015hqb,Mukhopadhyay:2016fre}. This also means the bulk dynamics is encoded in the RG of the boundary theory, and perhaps the bulk theory, and more importantly its gravitational dynamics, can be obtained from the RG of the boundary theory \cite{Douglas:2010rc,Lee:2009ij,Lee:2012xba,Radicevic:2011py,Swingle:2012wq,Sachs:2013pca,Mintun:2014gua,Lunts:2015yua,Kim:2020nst,Guijosa:2022jdo}.

In \cite{sathiapalan_holographic_2017}, a prescription has been laid out to obtain a $D+1$ dimensional dual bulk action from  a $D$-dimensional boundary theory using Polchinski's Exact Renormalisation Group (ERG) equation. In subsequent works, specialising to the $O(N)$ model  \cite{Sathiapalan:2019zex,Sathiapalan:2020cff}, the bulk scalar action upto cubic order, and the leading order in $\sqn$ action for the spin 1 and spin 2 fields \cite{dharanipragada_bulk_2022} has been obtained. In \cite{Dharanipragada:2023mkc}, the cubic coupling between the graviton and two scalars also has been obtained.

The bulk dual of the free $O(N)$ model is known to be the minimal bosonic Vasiliev theory in AdS$_4$ \cite{Klebanov:2002ja}. This bulk theory consists of fields of all even spins, and obey a higher spin algebra. These are dual to the conserved singlet currents of the $O(N)$ model, of which also there's one corresponding to each even spin, and they obey the same higher spin algebra. Vasiliev's theory is defined in terms of highly non-linear equations of motion with a high degree of redundancy in terms of auxiliary fields  \cite{Vasiliev:1999ba}. Thus, obtaining an action for the theory is of use.

There are several works attempting to define the Vasiliev theory through the 3D free $O(N)$ model \cite{Sleight:2016dba,Giombi:2009wh,Das:2003vw,deMelloKoch:2010wdf,Anninos:2011ui,Sachs:2013pca,deMelloKoch:2014vnt,Mintun:2014gua,Jin:2015aba,Aharony:2020omh,Aharony:2021ovo,Johnson:2022cbe,Solberg:2023akd}. The higher spins of the theory are all characterised by gauge freedom, which must play a role in their interactions. Therefore, in this paper we work with the simplest gauge field in the bulk, setting the rest to zero, and obtain its action upto cubic interaction from the ERG of the boundary theory's spin 1 current.

The $O(N)$ model contains only even spin currents. To look at odd spins, one must work with the $U(N)$ model. But the 3 point function of the spin-1 current in the $U(N)$ model is zero, and hence the bulk theory doesn't have a cubic interaction for this theory. We look at the simplest non-abelian case. When the boundary theory is the free $U(2N)$ vector model, and we consider the currents which are singlets of $USp(2N)$, the theory has conserved currents of all spins, one per even spin, and three per odd spin. The odd spin triplets are each in the adjoint representation of $\mathfrak{su}(2)$. 

\subsection{The prescription}
We obtain the bulk AdS theory from the boundary theory in the following steps.
\begin{enumerate}
	\item Introduce the singlet current of interest as an auxiliary field $\sigma_s$ in the action. (The subscript s stands for all the indices the current may have.) This will need us to introduce another auxiliary field $\chi^s$ which is a Lagrange multiplier enforcing the condition that $\sigma_s$ is the required current. 
	
	\item Integrate out the low energy modes of the fundamental vector field $\phi^I_l$ and obtain a low energy action depending on $\phi^I_l, \sigma_s, \chi^s$ and any sources. Set the high energy field $\phi_h^I=0$ and integrate out $\chi^s$ as well. The resulting action would be a low energy action for the $\sigma_s$ current. It gives the correlators for the current at low energies. We derive the ERG equation for the current from the Polchinski's ERG equation for the fundamental field.
	
	\item The ERG equation is analogous to the heat equation and we can obtain the ERG evolution operator as a path integral. The fields in this path integral depend also on $t$ (the scaling direction), $t=-\log (\Lambda/\Lambda_b)$, where $\Lambda_{b}$ is the bare momentum cut-off and $\Lambda$ is a generic cut-off. Therefore this is now a $D+1$ dimensional path integral.
	
	\item The $D+1$ dimensional action from this path integral is non-local. We change the $t$ to the  Poincar\'{e} $z$ coordinate by defining $z=\frac{1}{\Lambda_{b}} \e{t}=\frac{1}{\Lambda}$. We further redefine the field $\sigma_s$ which now depends on the bulk radial direction $z$ along with the boundary directions because of its scale dependence. $\sigma_s=fy_s$. $f$ is a smooth field redefinition which we define to cancel the non-local factor in the kinetic term. We further require $f$ to obey a differential equation that makes the quadratic term for $y_s$ the standard action in AdS$_{D+1}$ for a scalar field of mass $m^2=D(D-\Delta)$. This set of scalar fields can be identifed with a gauge-fixed spin-$n$ field on defining $y_s=z^n a_s$.
	This quadratic action is obtained as the quadratic part of  a gauge invariant kinetic term after gauge fixing.
	\item The $D+1$ dimensional path integral action contains cubic and higher order terms in $\sigma_s$. After the mapping $\sigma_s=fy_s$ they become polynomials in $y_s$. The question addressed in this paper then is whether the cubic terms (after the mapping) are local and also part of a gauge invariant action. It is shown that it is part of a local gauge invariant action.
	\item The differential equation imposed on $f$ is solved by modified Bessel functions. This determines the regulating function in the propagator for the fundamental field we use in the ERG. This complicated propagator shows up in the cubic and higher order terms in loop integrals. For the cubic term, this loop integral is difficult to tackle directly, so we replace the complicated regulator with a simpler regulator and perform the integral. This modification does not affect the boundary correlators computed by the bulk action. Remarkably, the result of the simpler integral cancels the Bessel functions appearing in the cubic term except for some $|p|$ like factors, where $p$ is the momentum of one of the fields.
	\item We systematically exchange these $|p|$ factors for $z$-derivatives and $p^2$ factors using the classical form of the field $a_s$ to obtain an action that we show to be a gauge-fixed action made from the gauge invariant field strength tensor $F_{MN}$. 
\end{enumerate}
Thus our final outcome is an action with a Yang-Mills term as well as an $\Tr (F^L_MF^M_NF^N_L)$ term. This action correctly reproduces the boundary correlators for the O(N) model and the correlators match the form required by \cite{Bzowski:2013sza} to obey the conformal Ward identities.

\section{The action}
We begin with the $U(2N)$ action in $D$ dimensions. Later we set $D=3$.
\begin{equation}
	S_B=\sqrt{N}\int_x \partial_\mu\varphi^*_A\pMU\varphi^A,
\end{equation}
and after separating the $USp(2N)$ and the $SU(2)$ indices the action becomes (see appendix \ref{current} for details)
\begin{equation}
	S=\frac12\sqrt{N}\int_x  \partial_\mu \phi^{Ia}\pMU \phi^{Jb} Q^{IJ}\epsilon_{ab},
\end{equation}
where $1\leq I,J\leq 2N$ are $USp(2N)$ indices and $a,b$ are the $j=1/2$ $SU(2)$ indices and
\begin{equation}
	Q\equiv \begin{pmatrix}
		0 & \mathds{1}_N\\
		-\mathds{1}_N & 0 
	\end{pmatrix}.
\end{equation}

We need to introduce auxiliary fields to stand for the currents, but the currents will run under RG, so we first begin by introducing sources for the currents, $A_\mu^i$. We introduce this background gauge field through covariantising the derivative.
\begin{equation}
	S_B=\frac12\sqrt{N}\int_x \Dmu \phi^{Ia} \DMU\phi^{Jb}\skewsym \levciv,
\end{equation}
where
\begin{equation}
	\Dmu \field\coloneqq \partial_\mu\field +A^i_\mu (T_i)^a_b \phi^{Ib},
\end{equation}
and $T_i$ are the generators of the $j=1/2$ representation of $\mathfrak{su}(2)$. The index $i$ is the $\mathfrak{su}(2)$ adjoint index.

We introduce the auxiliary field as functional derivative with respect to the gauge field.
\begin{equation}
	Z[A]=\int \CD\sigma \delta\bigg(\smd-\sqn\fd{\AMU_i}\bigg)Z[A].
\end{equation}
We also introduce a Lagrange multiplier field to enforce the delta functional.
\begin{align}
	Z[A] = &\int\CD\chi\CD\sigma\CD\phi \exp\Big\{i\int_x\chi^\mu_i\Big(\smd-\sqn\fd{\AMU_i}\Big)\Big\}\exp{\Big\{-S_B[\phi,\Amu]\Big\}}\nonumber\\
	=&\int\CD\chi\CD\sigma\CD\phi \exp\Big\{i\int_x\chi^\mu_i\smd\Big\}\exp\{-S_B[\phi,\Amu-\sqn i\cmd[]]\}\nonumber\\
	=&\int\CD\chi\CD\sigma\CD\phi \exp\Big\{\int_x(i\chi^\mu_i+\sqrt N \AMU_i)\smd \Big\}\exp\{-S_B[\phi,-\sqn i\cmd[]]\}.
\end{align}
In the last step, we have shifted $\chi_\mu$ by the field $A_\mu$. 

\section{ERG equation for the current}
Polchinski's Exact Renormalisation Group equation \cite{Polchinski:1983gv} (see also \cite{Wilson:1973jj,Wegner:1972ih,Wilson:1983xri,Wetterich:1992yh,Morris:1993qb,Bagnuls:2000ae,Bagnuls:2001pr,Igarashi:2009tj,Rosten:2010vm}) gives the rate of change of the interacting part of an action with its cutoff. But, here we work with the \textit{flipped} ERG equation\footnote{The flipped ERG equation is the equation obtained for flow of action of a composite field, (the current in this case), by integrating out more and more of the fundamental field \textit{from energy scale $0$ to $\Lambda$}. Consequently, using this action of the composite field, the propagator lines get contributions from fundamental modes between energy scales $0$ and $\Lambda$. This is the expected behaviour for an effective action for the composite field. But. as is clear from the derivation of the flipped ERG equation in appendix \ref{CurrentERGDerivation}, the contributions from fundamental modes above $\Lambda$ are set to zero. Thus, the flow for the composite field action is not the usual Wilsonian ERG. There is information lost, which is recovered as $\Lambda\to\infty$, or from bulk perspective,as we move close to boundary. See appendix \ref{solutionFlippedERG} and \cite{Dharanipragada:2023mkc} for more discussion.}, which occurs more naturally in Holographic RG contexts \cite{Faulkner:2010jy}. The action is split into two, one containing terms with low energy modes, which we integrate out and the other involving quadratic high energy terms. 
\begin{align}
	S_l[\phi_l,\phi_h,\cmd[]] = &\Qeps\int_x\Big\{ \frac12\sqrt{N}\fil\dli \fil[Jb] - i\cmu \gen(\fil[Ic]\partial_\mu \fil[Jb] + \fih[Ic]\partial_\mu\fil[Jb]+\fil[Ic]\partial_\mu\fih[Jb])\nonumber\\
	& - \frac12\sqn\gen \gen[j][b][d]\cmu\cmd[j](\fil[Ic]\fil[Jd] + 2\fih[Ic]\fil[Jd] ) \Big\}
%	=& Q^{IJ}\int_x\Big\{ \frac12\sqrt{N}\fil\dli\epsilon_{ab}\fil[Jb] - i\epsilon_{ab}\cmu \gen(\fil[Ic]\partial_\mu\fil[Jb] + 2 \fih[Ic]\partial_\mu\fil[Jb])\nonumber\\
%	& - \frac12\sqn\epsilon_{ab}\gen \gen[j][b][d]\cmu\cmd[j](\fil[Ic]\fil[Jd] + 2\fih[Ic]\fil[Jd] ) \Big\}
\end{align}
and
\begin{equation}
	S_h[\phi_l,\cmd[]] = \Qeps\int_x\Big\{ \frac12\sqrt{N}\fih[]\dhi\fih[] - i\cmu\gen \fih[Ic]\partial_\mu\fih[Jb] - \frac12\sqn\gen \gen[j][b][d]\cmu\cmd[j]\fih[Ic]\fih[Jd] \Big\},
\end{equation}
i.e.,
\begin{equation*}
	S_B = S_h + S_l,\quad \phi=\phi_h+\phi_l, \textrm{ and the propagator } \Delta = \Delta_h + \Delta_l.
\end{equation*}
$S_B$, i.e., the bare theory is defined at a cutoff $\Lambda_b$. We split the fields in this theory into two modes, with $\Delta_h$ only propagating modes between cutoff $\Lambda$ and bare cutoff $\Lambda_b$, and $\Delta_l$ only propagating modes below $\Lambda$.

The flipped ERG involves integrating out $\fil$, then using the ERG for $\fih$ to derive an ERG for $\smu$ by setting $\fih$ to zero and integrating out $\cmu$. We do this in the appendix \ref{CurrentERGDerivation} to obtain the ERG for the $\smu$ action.
%%%%%%%%%%%%%%%%%%%%%%%%%%%%%%%%%%%%%%%%%%%%%%%%%%%%%%%%%%%%%%%%%%%%%%%%%%%%%%%
 
 \section{Holographic Action}
 The flipped ERG equation \eqref{ERG} is of the form 
 \begin{equation}
 \ddt \psi = \Big(\frac12\dot{\mathcal{G}}\frac{\partial^2\hfill}{\partial x^2}+\sqn V(x)\Big)\psi,
 \end{equation}
 where $\mathcal{G} = I(p^2)$, given in \eqref{Isigmapropagator}. The evolution operator is then given in analogy with the Feynman Path Integral solution to the Schrödinger equation:
 \begin{align}
 \psi(x_f,T) =& \int \mathrm{d}x_i\int\limits_{x(0)=x_i}^{x(T)=x_f}\mathcal{D}x(t) \exp\bigg\{\int_{0}^{T}\mathrm dt \Big[-\frac{\dot{x}^2}{2\dot{\mathcal{G}}} + \sqn V(x(t))\Big] \bigg\} \psi(x_i,0)
 \label{analogypathintegral}\\
 \equiv& \int \mathrm{d}x_i U(x_f,T;x_i,0) \psi(x_i,0),
 \end{align}
 where $U$ is the evolution operator. 

The evolution operator takes $S_\Lambda[\smu]$ from $\Lambda=0$, where there it is $0$, to $\Lambda=\Lambda_b$ where it is the full action for $\sigma$.

 For the present case, we have the evolution operator as
 \begin{align}
 U[\sigma_f,t_f;&\sigma_i,t_i]= \int\limits_{\sigma(t_i)=\sigma_i}^{\sigma(t_f)=\sigma_f}\mathcal D \sigma^\mu  \exp\bigg\{ \int \mathrm dt \Big[-\frac12 \int_p \frac 1{\dot{I}(p^2)}\dsmu\dsmd \\
 &+\frac{2}{\sqrt N}\eps_{ijk}\int_p\int_q\int_r \dpqr\int_k\frac{\Dl[,k+p]\Dl[,k-q]\dot{\Delta}_{l,k}}{I(p^2)I(q^2)I(r^2)} k_{\mu}k_{\nu}(k+p)_{\rho}\smu(p)\snu[j](q)\sru(r)  \Big] \bigg\}.\nonumber
 \end{align}
 We interpret this as the functial integral for a $D+1$ dimensional field theory, with $t$ interpreted as an extra spacelike coordinate. We can read off the action for this theory from this functional integral.
 This ``bulk" action is given by 
 \begin{equation}
 	\boxed{
 		\begin{aligned}
 S[\sigma^{\mu}] = & \int \mathrm dt \Big[\frac12 \int_p \frac 1{\dot{I}(p^2)}\dsmu(p)\dsmd(-p) \label{bulkaction}\\
 &-\frac{2}{\sqrt N}\eps_{ijk}\int_p\int_q\int_r \dpqr\int_k\frac{\Dl[,k+p]\Dl[,k-q]\dot{\Delta}_{l,k}}{I(p^2)I(q^2)I(r^2)} k_{\mu}k_{\nu}(k+p)_{\rho}\smu(p)\snu[j](q)\sru(r)  \Big].
\end{aligned}
}
\end{equation}

%%%%%%%%%%%%%%%%%%%%%%%%%%%%%%%%%%%%%%%%%%%%%%%%%%%%%%%%%%%%%%%%%%%%%%%%%%%%%%%

\section{Mapping to AdS}
\label{mapToAdS}
The action obtained above \eqref{bulkaction} is nonlocal. With a field redefinition, we can bring the quadratic term into the standard AdS kinetic form.
\subsection{The field redefinition}
\label{fieldredefinition}
We suppress the $\mathfrak{su}(2)$ adjoint indices $i,j,k$ in the following.
The field redefinition function $f$ differs a bit from  \cite{dharanipragada_bulk_2022} for the kinetic term.
\begin{equation}
	f(p,t)= \sqrt{-\dot {I}(p^2, t)e^{Dt}}.
\end{equation}
We define
\begin{align}
	z&\equiv \frac{1}{\Lambda_b} e^{-t}=\frac{1}{\Lambda},\\
	f(p,z) &= \sqrt{z^{-D+1}\pz {I}(p^2,z)}\\
	\sigma_\mu(p,z)&\equiv zf(p,z)a_\mu(p,z)
	\label{mapping}
\end{align}
If we choose an $f(z,p)$ that solves the following differential equation, then the kinetic term of the action can be put in standard AdS form.
\begin{equation}
	\ddz\Big(z^{-D+1}\frac{\partial}{\partial z}\frac 1f\Big) = z^{-D+1}(p^2z^2+m^2)\frac 1f,
	\label{mapeq}
\end{equation}
where $m^2=1-D$ for the vector.

The kinetic term in the action then looks like 
\begin{equation}
 S_{K} = \frac12\int_p\int \frac{\mathrm dz}{z^{D-3}} (\pz \amup(p,z) \pz \amd(-p,z) +p^2 \amup(p,z)  \amd(-p,z)).
\end{equation}

The general solution to equation \eqref{mapeq} that satisfies the constraints of analyticity is described in \cite{dharanipragada_bulk_2022}. 
\begin{equation}
1/f=A(p) z^{\Dt} K_\nu (pz) + B(p) z^{\Dt} I_\nu(pz),
\label{fsolution}
\end{equation}
where  $I_\nu(x)$ and $K_\nu(x)$ are modified Bessel functions.
And as described in \cite{sathiapalan_holographic_2017}, $y_\mu=za_\mu$ satisfies the same differential equation as EOM. From \eqref{mapping} and \eqref{EOM}, thus, $I/f$ must also satisfy the same differential equation, and must be given in terms of the modified Bessel functions as $I/f=z^{\frac D2}(C(p) K_\nu (pz) + D(p) I_\nu(pz))$, for some $C(p), D(p)$. (That $I/f$ must satisfy the same differential equation can also be seen by substituting $-z^{D-1}f^2$ for $\partial_zI$ in the differential equation, on which it reduces to \eqref{mapeq}.) Hence, 
\begin{equation}
	I(p,z)=\frac{C(p) K_\nu (pz) + D(p) I_\nu(pz)}{A(p) K_\nu (pz) + B(p) I_\nu(pz)}.
	\label{Isolution}
\end{equation}
Further, 
\begin{equation}
	AD-BC=1,
	\label{wronskian}
\end{equation}
which results from taking the Wronskian of $I/f$ and $1/f$, which are two independent solutions of the differential equation \eqref{mapeq}.

The functions $A, B, C, D$ are fixed by the boundary behaviours of $f$ and $I$.
\begin{equation}
	I(p,0)\equiv -\gamma p^{2\nu}\mathrel{\mathop{=}\limits_{D=3}}-\gamma p; \text{ and } I(p,\infty)= 0,
	\label{Ibdry}
\end{equation}
where $\gamma p$ is the full Green's function at the boundary in $D=3$, and where $\nu^2=m^2+D^2/4$, which comes from solving the differential equation \eqref{mapeq} \cite{erbin_scalar_nodate}, with $m^2=1-D$ for the vector. We choose the positive root 
\begin{equation}
\nu=\frac D2-1=\frac32-1=\frac12.
\end{equation} 

The boundary behaviour of $f$ is determined as follows. Near the boundary, the leading behaviour of the bulk AdS field $y_\mu$ is given by 
\begin{equation}
y_\mu(x,z)\approx z^{D-\Delta}A_\mu(x),
\label{ybdry}
\end{equation}
where $A_\mu$ is the boundary source for the current $\sigma_\mu(x)$. $\Delta=\frac D2+\nu=2$ is the scaling dimension of $\sigma_\mu(x)$. The EOM for the bulk field $\sigma_\mu(z,x)$ is given by \eqref{EOM}
\begin{equation}
	\sigma_\mu(p,z)=b_\mu(p)I(p,z),
	\label{bulkEOM}
\end{equation}
for some $b_\mu(p)$. The boundary action is given by, (refer \eqref{o1waction}),
\begin{equation}
S_b[\sigma_{b\mu}]=\frac12\int_p\frac1{I_b(p)}\sigma_{b\mu}(p)\sigma^\mu_b(-p)+\int_pJ_\mu(p)\sigma^\mu_b(-p),
\end{equation}
where the subscript $b$ indicates $\sigma_{b\mu}(p)\equiv\sigma_\mu(p,z=\epsilon)$. The EOM given by this action is
\begin{equation}
\sigma_{b\mu}(p)=-A_\mu(p)I(p).
\end{equation}
Therefore, $b^\mu$ from \eqref{bulkEOM} is the boundary source $-A^\mu$ and $\sigma^\mu(p,z)=-A_\mu(p)I (p,z)$. Then, from the mapping equation \eqref{mapping} and boundary behaviours of $y_\mu$ \eqref{ybdry} and $I$ \eqref{Ibdry}, we have
\begin{equation}
\lim\limits_{z\to\eps}f(p,z)\approx \gamma z^{-D/2+\nu}p^{2\nu}.
\label{fbdry}
\end{equation}

Now that we have $3$ additional constraints on $A,B,C$ and $D$ given by boundary behaviours of $I$ \eqref{Ibdry} and $f$ \eqref{fbdry}, along with the Wronskian \eqref{wronskian}, we can determine their values.

The asymptotic behaviour of the modified Bessel functions is
\begin{align}
I_\nu(x) &\mathrel{\mathop{\longrightarrow}\limits_{x\to\infty}} \frac{1}{\sqrt{2\pi x}}e^{x},\\
K_\nu(x) &\mathrel{\mathop{\longrightarrow}\limits_{x\to\infty}} \sqrt{\frac{\pi}{2x}}e^{-x},
\end{align}
$\therefore$ for the second boundary condition of $I$ to be satisfied,
\begin{equation}
	D(p)=0.
\end{equation}
Thus, \eqref{wronskian} becomes 
\begin{equation}
BC=-1.
\end{equation}
At the other limit, $pz\to0$,
\begin{align}
I_\nu(x)=&\sum_{k=0}^{\infty}\frac{1}{\Gamma(k+\nu+1)k!}\Big(\frac x2\Big)^{2k+\nu},\\
K_\nu(x)=&\hf \Big[\Gamma(\nu)\Big(\frac x2\Big)^{-\nu}\Big(1+\frac{x^2}{4(1-\nu)}+\ldots\Big) +\Gamma(-\nu)\Big(\frac x2\Big)^{\nu}\Big(1+\frac{x^2}{4(1+\nu)}+\ldots\Big)\Big],~ \nu\notin\mathbb{Z},
\end{align}
$\therefore$ we have from \eqref{fbdry} and \eqref{fsolution},
\begin{equation}
	A(p)=\frac{1}{\Gamma(\nu)}2^{1-\nu}p^{-\nu}.
\end{equation}
\eqref{Ibdry} gives
\begin{equation}
	C(p)/A(p) = -\gamma p^{2\nu}.
\end{equation}
Then we have 
\begin{equation}
	C(p)= -\frac{\gamma}{\Gamma(\nu)}2^{1-\nu}p^\nu;\quad B(p)=\frac1\gamma\Gamma(\nu)2^{-1+\nu}p^{-\nu}.
\end{equation}
Thus,
\begin{align}
	1/f=& p^{-\nu} z^{\Dt} \bigg(\frac{1}{\Gamma(\nu)}2^{1-\nu}K_\nu (pz) + \frac1\gamma\Gamma(\nu)2^{-1+\nu} I_\nu(pz)\bigg),\\
	I(p,z) =& -\frac{\gamma^2 p^{2\nu} K_\nu(pz)}{\gamma K_\nu (pz) +2^{2\nu-2}\Gamma(\nu)^2 I_\nu(pz)}.
	\label{regularisation}
\end{align}

%%%%%%%%%%%%%%%%%%%%%%%%%%%%%%%%%%%%%%%%%%%%%%%%%%%%%%%%%%%%%%%%%%%%%%%%%%%%%%%

\subsection{The cublic term}
\label{mapping-cubic}
The $O(\sqn)$ term is \eqref{bulkaction}, substituting $a_\mu$ for $\smd[]$,
\begin{align*}
 	S_3 = & -\frac{2}{\sqrt N}\eps_{ijk}\int \mathrm dt \int_p\int_q\int_r \dpqr\int_k\frac{\Dl[,k+p]\Dl[,k-q]\dot{\Delta}_{l,k}}{I(p^2)I(q^2)I(r^2)} k_{\mu}k_{\nu}(k+p)_{\rho}\smu(p)\snu[j](q)\sru(r)\\
 	=& -\frac{2}{\sqrt N}\eps_{ijk}\int \frac{\mathrm dz}{z} \int_p\int_q\int_r \dpqr\int_k\frac{\Dl[,k+p]\Dl[,k-q]z\pz{\Delta}_{l,k}}{I(p^2)I(q^2)I(r^2)} k_{\mu}k_{\nu}(k+p)_{\rho} \\
 	&\times z^3f(p)f(q)f(r) \amup(p) a^{\nu}_j(q) a^{\rho}_k(r).
\end{align*}
Note here that only z-dependent part of this whole term is $\Dl[,k+p]\Dl[,k-q]\pz{\Delta}_{l,k}$. The rest is a $z$-independent considering \eqref{constantRatio}.

We can substitute from \eqref{regularisation}
\begin{equation}
	\frac{f(p,z)}{I(p,z)}=-\frac{2^{-1+\nu}\Gamma(\nu)z^{-D/2}}{\gamma p^{\nu}K_\nu(pz)}=-\sqrt{\frac \pi2}\frac{z^{-3/2}}{\gamma p^{\nu}K_\nu(pz)},
\end{equation}
giving
\begin{align}
	S_3 = &\frac{2 }{\sqrt N\gamma^3}\eps_{ijk}\bigg(\frac \pi2\bigg)^{3/2}\int z^{-3/2}\mathrm dz \int_p\int_q\int_r \dpqr\int_k\Dl[,k+p]\Dl[,k-q]\pz{\Delta}_{l,k} \nonumber\\ &\quad\times k_{\mu}k_{\nu}(k+p)_{\rho}\frac{p^{-\nu}}{K_\nu(pz)}\frac{q^{-\nu}}{K_\nu(qz)}\frac{r^{-\nu}}{K_\nu(rz)}
	\times \amup(p) a^{\nu}_j(q) a^{\rho}_k(r)\nonumber\\
	= &\frac{2}{3\sqrt N\gamma^3}\eps_{ijk}\bigg(\frac \pi2\bigg)^{3/2}\int z^{-3/2}\mathrm dz \int_p\int_q\int_r \dpqr\int_k\pz(\Dl[,k+p]\Dl[,k-q]{\Delta}_{l,k}) \nonumber\\ &\quad\times k_{\mu}k_{\nu}(k+p)_{\rho}\frac{p^{-\nu}}{K_\nu(pz)}\frac{q^{-\nu}}{K_\nu(qz)}\frac{r^{-\nu}}{K_\nu(rz)}
	\times \amup(p) a^{\nu}_j(q) a^{\rho}_k(r).
\end{align}
As mentioned before, the quantities outside $\pz$ parentheses are $z$-independent, hence the integrand is a total derivative in $z$. Thus we could evaluate it at the limits of the integral. We can make use of this fact to simplify the evaluation of the integral. 
The low energy propagator is determined by the RG regularisation chosen\footnote{In \cite{dharanipragada_bulk_2022}, we discussed that when we map the ERG equation to an AdS bulk action, it constrains the regulator we can use for the ERG equation. Since this must be done for each operator that deforms the boundary CFT, there are conflicting constraints on the high energy propagator. This was resolved in \cite{dharanipragada_bulk_2022} and the constraint imposed by the ERG equation of the scalar composite operator discussed in \cite{Sathiapalan:2020cff} was chosen. We follow the same choice here.}. But this is a complicated propagator. We instead choose to work with a simple low energy propagator that is more tractable. The error in correlators calculated with this is $O(p/\Lambda_b)$. Thus, in the limit of $\eps\to0$, or $\Lambda_b\to\infty$, we get the correct correlators. The $k$ integral is done in appendix \ref{integral}. 

The result is \eqref{localResultOfIntegral}.
\begin{align}
\int_k\pz(\Dl[,k+p]\Dl[,k-q]{\Delta}_{l,k}) k_{\mu}k_{\nu}(k+p)_{\rho}\approx
	\frac{1}{16(2\pi)^{5/2}} z^{3}\Big[-q_\mu r_\nu p_\rho z^2-\dmn p_\rho(1+{rz})
	-q_\mu \delta_{\nu\rho}(1+{pz})\nonumber\\
	-r_\nu \delta_{\mu\rho} (1+{qz})\Big]
	\Big(\frac{r}{z}\Big)^{\nu}\Big(\frac{p}{z}\Big)^{\nu}\Big(\frac{q}{z}\Big)^{\nu}K_{\nu}(rz)K_{\nu}(pz)K_{\nu}(qz).
\end{align}
The interaction term is then
\begin{align}
S_3  =  &\frac{1}{\sqrt N}\frac{1}{384\pi\gamma^3}\eps_{ijk}\int\mathrm dz \int_p\int_q\int_r \dpqr\Big[-z^2q_\mu r_\nu p_\rho-\dmn p_\rho(1+{rz})
-q_\mu \delta_{\nu\rho}(1+{pz})\nonumber\\
&\qquad-r_\nu \delta_{\mu\rho} (1+{qz})\Big]
\times \amup(p) a^{\nu}_j(q) a^{\rho}_k(r)\nonumber\\
=&\frac{1}{\sqrt N}\frac{1}{384\pi\gamma^3}\eps_{ijk}\int\mathrm dz \int_p\int_q\int_r \dpqr [-z^2q_\mu r_\nu p_\rho\amup(p) a^{\nu}_j(q) a^{\rho}_k(r)\nonumber\\
&\qquad-3(1+{rz})
 \amup(p) a_{\mu}^j(q)\times p_\rho a^{\rho}_k(r)]
 \label{cubicTerm}
\end{align}

This action looks unfamiliar and has no gauge invariance manifest. But the gauge invariance is hidden.
In the next section, we show that this action is equivalent to the gauge fixed version of a manifestly gauge invariant action.

%%%%%%%%%%%%%%%%%%%%%%%%%%%%%%%%%%%%%%%%%%%%%%%%%%%%%%%%%%%%%%%%%%%%%%%%%%%%%%%

\section{Gauge invariance}
\label{GaugeInvariance}
Even though our cubic term is non-standard and the gauge invariance is not manifest, it is equivalent classically to the cubic term in a gauge invariant action. In this section we show that upon using the equations of motion, the following gauge invariant action reduces to \eqref{cubicTerm}.

\begin{equation}
	S'[a^\mu]= \frac14 \int \der^{D+1}x\sqrt{G}\,  F^{MNi}(x,z)F_{MN}^i(x,z) + g'\epsilon^{ijk}\int \der^{D+1}x\sqrt{G}\,  F^{Li}_{\ M}(x,z)F_{\ N}^{Mj}(x,z) F_{\ L}^{Nk}(x,z),
	\label{gaugeinvariantaction}
\end{equation}
where $L,M,N$ are $D+1$ dimensional indices that run over $\{x^\mu,z\}$; $x^\mu$ are the boundary coordinates.
\begin{equation}
	F_{MN}^i(x,z)=\partial_M a^i_N(x,z)-\partial_N a^i_M(x,z)- ig \eijk a^j_M(x,z)a^k_N(x,z),
\end{equation}
and $g$ and $g'$ are the YM coupling and the $FFF$ coupling respectively.
The metric is 
\begin{equation}
	\der s^2=\frac1{z^2}(dz^2+\dmn \der x^\mu \der x^\nu),
\end{equation}
therefore $\sqrt{G}=z^{-D-1}$.

We fix the gauge to be in ``radial" gauge, i.e.,
\begin{equation}
	a_z^i(x,z)=0,
\end{equation}
and we use the residual gauge freedom to fix
\begin{equation}
	\partial_\mu a^{\mu i}(x,z)=0.
\end{equation}
The fields obtained from ERG of the boundary theory are in this gauge. 

On fourier transforming the boundary coordinates, the cubic terms from this gauge fixed action are
\begin{align}
	S_3^{\prime YM} = \frac12 g\eijk \int\limits_{p,q,r} & \dpqr \sqG z^{4}(p_\mu a^i_\nu(p,z)-p_\nu a^i_\mu(p,z))a^{\mu j}(q,z)a^{\nu k}(r,z)\\
	S_3^{\prime FFF} =  -i g' \eijk \int\limits_{p,q,r} & \dpqr \sqG z^{6} [(p_\mu a^{\nu i}(p,z)-p^\nu a^i_\mu(p,z)) (q_\nu a^{\rho j}(q,z)-q^\rho a^j_\nu(q,z))\nonumber\\
	& (r_\rho a^{\mu k}(r,z)-r^\mu a^k_\rho(r,z))\nonumber\\
	&+ (p_\mu a^{\nu i}(p,z)-p^\nu a^i_\mu(p,z))\partial_z a_\nu^j(q,z)\partial_z a^{\mu k}(r,z\nonumber)\\
	&+ (q_\mu a^{\nu i}(q,z)-q^\nu a^i_\mu(q,z))\partial_z a_\nu^j(r,z)\partial_z a^{\mu k}(p,z)\nonumber\\
	&+ (r_\mu a^{\nu i}(r,z)-r^\nu a^i_\mu(r,z))\partial_z a_\nu^j(p,z)\partial_z a^{\mu k}(q,z)]
\end{align}
Since $p,q,r$ are dummy variables, the terms which are same upto permutations of these are the same, but we have written them out for clarity.

With some manipulation using $p^\mu+q^\mu+r^\mu=0$, and that the fields are transverse, we have
\begin{align}
	S'_3= -\eijk \int\limits_{p,q,r} &\dpqr \bigg[a_\mu^i(p,z)a_\nu^j(q,z)a_\rho^k(r,z)\Big(g\,  p^\rho\delta^{\mu\nu}+ig'z^2\Big(2p^\rho q^\mu r^\nu\nonumber\\
	 &-3(p\cdot r + r\cdot q)p^\rho\dumn \Big)\Big) +3ig' z^2 \, p^\rho \dumn \Big(a_\mu^i(p,z)\partial_z a_\nu^j(q,z)\partial_z a_\rho^k(r,z) \nonumber\\
	 &+\partial_z a_\mu^i(p,z) a_\nu^j(q,z)\partial_z a_\rho^k(r,z)\Big)\bigg]\\
	 =-\eijk \int\limits_{p,q,r} &\dpqr \bigg[a_\mu^i(p,z)a_\nu^j(q,z)a_\rho^k(r,z)\Big(g\,  p^\rho\delta^{\mu\nu}+ig'z^2\Big(2p^\rho q^\mu r^\nu\nonumber\\
	 &+3r^2\, p^\rho\dumn \Big)\Big) +3ig'z^2 \,  p^\rho \dumn \partial_z\Big(a_\mu^i(p,z) a_\nu^j(q,z)\Big)\partial_z a_\rho^k(r,z)\bigg].
	 \label{GIcubicterm}
\end{align} 

Now we show that \textit{on-shell}, the action we have obtained from ERG in section \ref{mapToAdS}, in \eqref{cubicTerm} is equivalent to this expression resulting from gauge-fixing a gauge invariant action. The evaluation of correlators is done using the large $N$ semi classical limit and so using the on-shell form of the action is not an additional restriction.

In \ref{fieldredefinition}, we have shown that the redefined bulk field $a^\mu$ is given by
\begin{equation}
	a_\mu^i(p,z)=\gamma \frac{2^{1-\nu}}{\Gamma(\nu)}A_\mu^i(p)z^{\frac D2-1}	|p|^\nu K_\nu(pz),
	\label{bulksolution}
\end{equation} 
where $A_\mu^i(p)$ is the source for the boundary current. Now for the boundary current, $\nu=\frac D2-1=\frac 12$, and $K_{1/2}(x)=\sqrt\frac{\pi}{2x}\e{-x}$. Thus $a^i_\mu(p,z)$ only depends on $z$ via $\e{-pz}$, and we have
\begin{equation}
	\pz a^i_\mu(p,z)=-p a^i_\mu(p,z).
\end{equation}
This means that we could exchange factors of $p$ in the action for $\pz$ acting on the gauge fields. Then the ERG derived action of \eqref{cubicTerm} becomes
\begin{align}
	S_3  
	=&\frac{1}{\sqrt N}\frac{1}{384\pi\gamma^3}\eps_{ijk}\int\mathrm dz \int_p\int_q\int_r \dpqr [-z^2q_\mu r_\nu p_\rho\amup(p,z) a^{\nu}_j(q,z) a^{\rho}_k(r,z)\nonumber\\
	&\qquad-3
	\amup(p,z) a_{\mu}^j(q,z)\times p_\rho a^{\rho}_k(r,z) +3z \amup(p,z) a_{\mu}^j(q,z)\times p_\rho \pz a^{\rho}_k(r,z)]
\end{align}

%The equations of motion to leading order in $g,g'$ are
%\begin{equation}
%	\partial_M F^{M i}=0,
%\end{equation}
%where we keep only $O(g^0)$ terms in the equation. In our chosen gauge, and moving to the boundary metric $\dmn$, this becomes
%\begin{equation}
%	\partial^2 a^{\nu i} +\partial_z^2 a^{\nu}=0.
%\end{equation}
%Fourier-transforming in the transverse directions,
%\begin{equation}
%	(-p^2 + \partial_z^2)a^{\nu i}(p,z)=0.
%	\label{GIeom}
%\end{equation}
This matches the action in \eqref{GIcubicterm} except for the third term, $3z \amup(p,z) a_{\mu}^j(q,z)\times p_\rho \pz a^{\rho}_k(r,z)$. We add the surface term $-\frac32\pz(z^2\amup(p,z) a_{\mu}^j(q,z)\times p_\rho \pz a^{\rho}_k(r,z))$, which goes to $0$ as $z\to 0$, and we're left with
\begin{equation}
	-\frac32z^2\pz(\amup(p,z) a_{\mu}^j(q,z))\times p_\rho \pz a^{\rho}_k(r,z) -\frac32z^2\amup(p,z) a_{\mu}^j(q,z)\times p_\rho \pz^2 a^{\rho}_k(r,z).
\end{equation}
Finally, we exchange the two derivatives in the second term for $r^2$ to get
\begin{equation}
	-\frac32z^2\pz(\amup(p,z) a_{\mu}^j(q,z))\times p_\rho \pz a^{\rho}_k(r,z) -\frac32p^2z^2\amup(p,z) a_{\mu}^j(q,z)\times p_\rho a^{\rho}_k(r,z).
\end{equation}
%Thus, using this EOM, \eqref{GIcubicterm} becomes
%\begin{align}
%		S_3
%	=-\eijk \int\limits_{p,q,r} &\dpqr \bigg[a_\mu^i(p,z)a_\nu^j(q,z)a_\rho^k(r,z)\Big(g\,  p^\rho\delta^{\mu\nu}+2ig'z^2 p^\rho q^\mu r^\nu\Big)\nonumber\\
%	& +3ig'z^2 \,  p^\rho \dumn \partial_z\Big(a_\mu^i(p,z) a_\nu^j(q,z)\partial_z a_\rho^k(r,z)\Big)\bigg]\\
%	=-\eijk \int\limits_{p,q,r} &\dpqr \bigg[a_\mu^i(p,z)a_\nu^j(q,z)a_\rho^k(r,z)\Big(g\,  p^\rho\delta^{\mu\nu}+2ig'z^2 p^\rho q^\mu r^\nu\Big)\nonumber\\
%	& -6ig'z \,  p^\rho \dumn a_\mu^i(p,z) a_\nu^j(q,z)\partial_z a_\rho^k(r,z)\bigg],
%\end{align}
%upto a surface term whose on-shell contribution vanishes at $z\to0$ boundary.
%
%Further, the EOM \eqref{GIeom} means $a_\nu^i(p,z)$ only depends on $z$ as $\e{\pm |p|z}$. But, $\e{|p|z}$ is ruled out because it is not well-behaved in the interior. Therefore, $a_\nu^i(p,z)\propto \e{|p|z}$. Therefore, we have $\partial_z a^{\nu i}(p,z)=|p| a^{\nu i}(p,z)$, and the action can be written as
%\begin{equation}
%	S_3
%	=-\eijk \int\limits_{p,q,r} \dpqr a_\mu^i(p,z)a_\nu^j(q,z)a_\rho^k(r,z)\Big((g-6i g' |r|z)\,  p^\rho\delta^{\mu\nu}+2ig'z^2 p^\rho q^\mu r^\nu\Big)
%\end{equation}

Comparing this with \eqref{GIcubicterm}, we see that they both match with $g$ and $g'$ given by
\begin{equation}
	g=\sqn \frac{1}{128\pi \gamma^3};\qquad g'=\frac {ig}6.
\end{equation}
%Firstly, the usual YM action in ``D" dimensional momentum space is 
%\begin{equation}
%	 \int\frac{\der^D p}{(2\pi)^D} F_{MN}^{i}(p)F^{MNi}(-p),
%\end{equation}
%where $M,N$ are the spacetime indices and $i$ is the gauge group adjoint index. In the present case we are interested in the YM action in AdS spacetime, with the $D$ flat directions Fourier transformed and the $z$ direction left as it is.
%\begin{equation}
%	\int \der z\int \frac{\der^D p}{(2\pi)^D} F_{MN}^{i}(p,z)F^{MNi}(-p,z),
%	\label{standardYM}
%\end{equation}
%The field strength is given by
%\begin{align}
%	F_{\mu\nu}^i&=p_\mu A^i_\nu(p,z)-p_\nu A^i_\mu(p,z)+ig \epsilon_{ijk}\int  \frac{\der^D q}{(2\pi)^D} A^j_\mu(q,z)A^k_\nu(p-q,z),\\
%	F_{\mu z}^i&=p_\mu A^i_z(p,z)-\partial_z A^i_\mu(p,z)+ig \epsilon_{ijk}\int  \frac{\der^D q}{(2\pi)^D} A^j_\mu(q,z)A^k_z(p-q,z).
%\end{align}
%Here we have taken the gauge group to be ${su}(2)$
%This is invariant under the gauge transformation
%\begin{align}
%	A^i_\mu(p,z)&\to A^i_\mu(p,z)+p_\mu \Lambda^i(p,z)-ig \epsilon_{ijk}\int  \frac{\der^D q}{(2\pi)^D} \Lambda^j(q,z)A^k_\mu(p-q,z),\\
%	A^i_z(p,z)&\to A^i_z(p,z)+\partial_z\Lambda^i(p,z)-ig \epsilon_{ijk}\int  \frac{\der^D q}{(2\pi)^D} \Lambda^j(q,z)A^k_z(p-q,z).
%\end{align}

%%%%%%%%%%%%%%%%%%%%%%%%%%%%%%%%%%%%%%%%%%%%%%%%%%%%%%%%%%%%%%%%%%%%%%%%%%%%%%%

\section{Summary and Conclusions}
One of the intriguing issues in holography is the emergence of gauge symmetry in the bulk theory. This is definitely required for consistency of the bulk theory as a quantum field theory. If the ERG approach of  constructing the bulk action from first principles, described  in \cite{sathiapalan_holographic_2017}, is to be applicable to all such theories, it is imperative that the mechanism that ensures this gauge invariance be understood better. In earlier papers that followed this approach only the Abelian part had been explored \cite{Dharanipragada:2023mkc}.  The non Abelian structure was not explored. This paper is a step in that direction.

In this paper we  start from a boundary scalar field theory that possesses a global $SU(2)$ symmetry. The correlators of the corresponding currents have a well defined form fixed by the Ward Identities. We construct a dual bulk action in AdS space from first principles using the ERG approach described in \cite{sathiapalan_holographic_2017}.
The result is an action for a Yang Mills gauge field (in holographic gauge $A_z=0$) in the bulk. This action is shown to be the cubic part of the usual gauge invariant Yang-Mills action quadratic in field strengths along with a  term cubic in Yang Mills field strength . The action is given in \eqref{gaugeinvariantaction} and verifies the consistency of this method.

There are many open questions. One is to extend this to the cubic interaction in gravity and indeed for all massless higher spins. It would be interesting if this can be done in a way that makes manifest the gauge symmetry so that the resulting bulk action is guaranteed to have the required symmetry. 

The other issue is that of locality. It is not clear what notion of locality a ``good" bulk dual should have. As emphasized earlier, the ERG approach to AdS/CFT guarantees that the bulk dual reproduces all correlators of the boundary theory. It also is designed to give a local kinetic term in AdS space for the bulk fields. But the locality of cubic and higher order terms is not guaranteed. It was verfied for the cubic scalar self interaction  in \cite{Sathiapalan:2020cff} and and for the cubic minimal coupling of the gravitational perturbation to the scalar in \cite{Dharanipragada:2023mkc} (and in the present paper for the Yang-Mills gauge field self-coupling). A better understanding of this issue is certainly required.

 The higher-spin-vector-model duality has been the playground for constructing bulk from boundary theory in several works. Some of these use the (E)RG approach \cite{Sachs:2013pca,Mintun:2014gua,Jin:2015aba}, while others use collective fields or bilocal holography \cite{Das:2003vw,deMelloKoch:2010wdf,deMelloKoch:2014vnt,Aharony:2020omh,Aharony:2021ovo,Solberg:2023akd}. There is a need to understand the relation between these approaches and ours. \cite{Aharony:2021ovo} also deals with obtaining a gauge theory in the bulk, albeit abelian. The connection has been explored by \cite{Mintun:2014gua} along with the question of obtaining an RG cutoff that naturally corresponds to the bulk radial direction. We believe our approach described in the series of papers \cite{sathiapalan_holographic_2017,Sathiapalan:2019zex,Sathiapalan:2020cff,dharanipragada_bulk_2022,Dharanipragada:2022yxw,Dharanipragada:2023mkc}, and the present one, answers this issue of the RG cutoff. 

Finally there is the question of generalizing th ERG approach to other space times such as flat space \cite{Sathiapalan:2024tdz} and de Sitter \cite{Dharanipragada:2022yxw}.

 \section*{Acknowledgements}
 P. D. would like to thank Nemani V. Suryanarayana for useful discussions and for appendix \ref{current}. P. D. would also like to thank the Insitute of Mathematical Sciences where the bulk of this work was completed while he was a graduate student there.

\appendix
\section{The current}
\label{current}
We begin with the free $U(2N)$ invariant scalar model.
\begin{equation}
	S_B=\sqrt{N}\int_x \partial_\mu\varphi^*_A\pMU\varphi^A,
\end{equation}
where $A$ is the $U(2N)$ index. As long as we don't introduce any interactions, the action is also invariant under a larger group of transformations: $SO(4N)$, the group of rotations of the $4N$ real scalars that are the real and imaginary parts of the complex $\varphi^A$. $SO(4N)$ has a maximal subgroup $USp(2N)\times SU(2)$. Our objective is to separate the indices corresponding to these two subgroups, so that we may consider operators which are traced over the $USp(2N)$.
%We split the $2N$ indices into two sets of $i,j\in\{1,2,\ldots,N\}$ and $a,b\in\{1,2\}$, so that the fields $\varphi^{ia}$ now manifest the maximal subgroup $SU(N)\times SU(2)$ transformations.

Now define new fields $\phi^{Ia}$ in terms of $\varphi^{A}$ with $I\in \{1,2,\ldots,2N\}$ and $a\in\{1,2\}$ as follows. For $I=i\leq N$,
\begin{equation}
	\phi^{I1}\coloneqq \varphi^i; \quad \phi^{I2}\coloneqq\varphi^*_{N+i},
\end{equation}
and for $I=N+i$, $1\leq i\leq N$,
\begin{equation}
	\phi^{Ia}\coloneqq \epsilon^{ab}\phi^{i*}_{b}.
\end{equation}
$\epsilon_{ab}$ is the two dimensional Levi-Civita tensor and $\epsilon^{ab}$ is its inverse. $\epsilon_{12}=1$. That is,
\begin{equation}
	\phi^{I1}=\begin{pmatrix}
		\varphi^{i}\\ -\varphi^{N+i} 
	\end{pmatrix}
	\text{ and }
	\phi^{I2}=\begin{pmatrix}
		\varphi^*_{N+i}\\ \varphi^*_{i}.
	\end{pmatrix}
\end{equation}
The action is now
\begin{equation}
	S=\sqrt{N}\int_x \partial_\mu\varphi^*_A\pMU\varphi^A\equiv \frac12\sqrt{N}\int_x  \partial_\mu \phi^{Ia}\pMU \phi^{Jb} Q^{IJ}\epsilon_{ab},
\end{equation}
where 
\begin{equation}
	Q\equiv \begin{pmatrix}
		0 & \mathds{1}_N\\
		-\mathds{1}_N & 0 
	\end{pmatrix}.
\end{equation}
In relabelling the fields $\phi$ instead of $\varphi$, we have only shuffled around the $4N$ real fields that are the real and imaginary parts of the complex $\varphi^A$. So, the action must still be $SO(4N)$ invariant. But, in this avatar, the index structure makes manifest the invariance of the action under the maximal subgroup $USp(2N)\times SU(2)$ of $SO(4N)$. $I,J$ are $USp(2N)$ indices and $a,b$ are $SU(2)$ indices. Because, $USp(2N)$ is the set of unitary transformations of determinant $1$ that leaves the matrix $Q$ invariant, i.e., complex $2N\times 2N$ matrices $S$ such that $S Q S^T=Q$ and $\det S=1$. And, $SU(2)$ transformations leave $\epsilon$ invariant. 

The global $SO(4N)$ symmetry means there are conserved currents
\begin{equation}
	J_\mu^{IJab}\equiv \phi^{Ia}\partial_\mu \phi^{Jb},
\end{equation}
which are in the adjoint representation  of $SO(4N)$. We select the $USp(2N)$ singlets\footnote{We select the $USp(2N)$ singlets instead of the $SO(4N)$ singlet so that the bulk dual is a non-abelian gauge field \cite{Giombi:2013fka}.} among them by weakly gauging the $USp(2N)$ and then taking the zero coupling limit \cite{Giombi:2016ejx}. With the index structure we have chosen, it is easy to make the currents $USp(2N)$ invariant. We contract them with $Q_{IJ}$. We then obtain three currents in the adjoint representation of $SU(2)$.
\begin{align}
	J^\mu_-&\equiv Q^{IJ} \phi^{I1}\pMU \phi^{J1},\\
	J^\mu_3&\equiv \frac1{{2}} Q^{IJ}( \phi^{I1}\pMU \phi^{J2}+ \phi^{I2}\pMU \phi^{J1}),\\
	J^\mu_+&\equiv Q^{IJ} \phi^{I2}\pMU \phi^{J2}.
\end{align}
Written in terms of the original fields, these are
\begin{align}
	J^\mu_- & =  Q^{AB} \varphi^A \partial^{\mu} \varphi^B, \\
	J^\mu_3 & =  \frac{1}{{2}} (\varphi^{\ast}_A \partial^{\mu} \varphi^A -
	\varphi^A \partial^{\mu} \varphi^{\ast}_A), \\
	J^\mu_+ & =  Q_{AB} \varphi^{\ast}_A \partial^{\mu} \varphi^{\ast}_B . 
\end{align}

\section{Deriving the ERG for the current}
\label{CurrentERGDerivation}
To this end, we separate the quadratic and linear terms in the action. If no spacetime parameter is specified explicitly for any field, it is $x$.
\begin{align}
	S_l^{(2)} =& \Qeps\int_x\Big\{ \frac12\sqrt{N}\fil\dli\fil[Jb] - i\cmu \gen\fil[Ic]\partial_\mu\fil[Jb] - \frac12\sqn\gen \gen[j][b][d]\cmu\cmd[j]\fil[Ic]\fil[Jd] \Big\}\nonumber\\
	=& \frac12Q^{IJ} \int_x\int_y \fil \big[\sqrt{N}\dli(x-y) \epsilon_{ab} \nonumber\\
	&- \underbrace{\big(2i\epsilon_{cb}\cmu\gen[i][c][a]\pmu + \sqn\epsilon_{cd}\cmu[i]\cmd[j] \gen[i][c][a] \gen[j][d][b]\big)\dxy}_{\K}  \big] \fil[Jb](y)\nonumber\\
	\equiv& \frac12 \skewsym\int_x\int_y \fil \big[\sqrt{N}\dli\epsilon_{ab} - \K \big]_{xy} \fil[Jb](y).
\end{align}
We have defined $\K$, 
\begin{equation}
	\K(x,y) = \Big[i\cmu \{\levciv[cb]\genca+\levciv[ac]\gen[i][c][b]\}\pmu + \sqn\cmu[i]\cmd[j] \levciv[cd]\genca \gen[j][d][b]\Big]\delta(x-y).
	\label{Kab}
\end{equation}
Note that $\K=-\K[ba]$.

The linear terms:
\begin{equation}
	S_l^{(1)}=-\skewsym\int_x\int_y\fil\K(x,y)\fih[Jb](y).
\end{equation}
The path integral is then 
\begin{align}
	Z[A] =&  \int\CD\chi\CD\sigma\CD\phi_h \exp\Big\{\int_x(i\cmu+\sqrt N\AMU_i)\smd \Big\}\exp\{-S_h[\phi_h,\cmd[]]-\sil[\phi_h,\cmd[]]\}\\
	\equiv&	\int\CD\sigma\CD\phi_h \e{-S_{\Lambda}[\fih[],\smu[]]},
	\label{pathintegral}
\end{align}
with
\begin{align}
	\e{-\sil}\equiv& \int\CD\fil \exp\{-S_l^{(1)}-S_l^{(2)}\}\\
	=& \exp \Big\{-\frac 12\Tr\log\Big[\skewsym(\sqrt{N}\dli\epsilon_{ab} - \K)\Big] \nonumber\\
	&+ \frac12 \skewsym \int\limits_u\int\limits_v  \Big(\int\limits_x\K[ac](u,x)\fih(x)\Big) \Big[\sqrt{N}\dli - K\Big]^{-1cd}(u,v) \Big(\int\limits_y\K[db](v,y)\fih[Jb](y)\Big) \Big\}.
\end{align}
The subscript $\Lambda$ indicates the action has an IR cutoff $\Lambda$. Taking the limit $\Lambda\to\infty$ then implies the path integral is fully calculated. The subscript $I$ stands for ``interaction''.

The flipped Polchinski ERG equation governs how the quantity $S_\Lambda[\phi_h]$ changes with change in $\Lambda$. It is the counterpart of the Polchinski ERG equation. The rate of change is in terms of the ``ERG time" $t\equiv\log(\Lambda/\Lambda_b)$. $t=-\infty$ corresponds to theory which has not been integrated over, and $t=0$ corresponds to the complete theory being intergated out. The equation is
\begin{equation}
	\ddt \e{-\sil} = -\frac12 \skewsym \epsilon^{ab} \sqn \int_x\int_y \dot{\Delta}_{l}(x,y)\frac{\delta^2 e^{-\sil}}{\delta \fih\delta\fih[Jb](y)}.
	\label{ergphi}
\end{equation}
We will set $\fih$ to $0$ after obtaining this expression, so we will only have to evaluate the following.
\begin{equation}
	\frac{\delta^2 \sil}{\delta \fih\delta\fih[Jb](y)} = -\skewsym\int_u\int_v\K[ac](u,x)[\sqrt N\dli\epsilon - K]^{-1cd}(u,v) \K[db](v,y).
\end{equation}
%As an illustration, to know what the derivative in $K$ act on,
%\begin{align}
%	\int_x\int_y \dot{\Delta}_{hxy}\frac{\delta^2 e^{-\sil}}{\delta \fih\delta\fih(y)} = &4N\int_x\int_y\int_u\int_v \dot{\Delta}_{hxy}\cmu[ik](u)\frac{\partial}{\partial u^{\mu}}\delta(u-x)[\ldots]^{-1kl}_{uv}\cnu[il](v)\frac{\partial}{\partial v^{\nu}}\delta(v-y)\nonumber\\
%	=& 4N\int_u\int_v \cmu[ik](u)\cnu[il](v)[\ldots]^{-1kl}_{uv}\frac{\partial^2 \dot{\Delta}_{huv}}{\partial u^{\mu}\partial v^{\nu}}
%\end{align}
Expanding the inverse matrix in the centre in a power series,
\begin{align*}
	\frac{\delta^2 \sil}{\delta \fih\fih[Jb](y)} =& -\sqn \skewsym\epsilon^{ce} \int_u\int_v\int_{z} \K[ac](u,x)[1 - \sqn \Delta_l \epsilon K]^{-1d}_e({u,z}) \Delta_l(z ,v) \K[db](v,y)\\
	=& -\sqn\skewsym\epsilon^{ce} \int_u\int_v\int_z \K[ac](u,x) \bigg[\delta^d_e\delta(u-z) +\sqn\eps^{df}\int_w \Delta_l(u,w)\K[fe](w,z) \nonumber\\
	& +\frac 1N [\Dl[] \eps K\Dl[] \eps K]^{d}_{e}(u,z) + \ldots\bigg] \Delta_l(z,v) \K[db](v,y)\\
	=& -\sqn\skewsym\epsilon^{ce} \int_u\int_v \K[ac](u,x)\Dl[](u,v) \K[eb](v,y) \nonumber\\
	& -\frac1N\skewsym\epsilon^{ce}\eps^{df}\int_u\int_v \K[ac](u,x)\K[db](v,y)\int_z\int_w\Dl[](u,w) \K[fe](w,z)\Dl[](z,y) + O\Big(K^4\Big).
\end{align*}
We look at the terms on the R.H.S. in (\ref{ergphi}) in detail.

\paragraph{Term 1}
%In the following, and later, we associate the index $\mu$ with $x$, $\nu$ with $y$, and $\rho$ with $z$.
The following are the terms resulting from the $K^2$ term. Note that $\skewsym\skewsym=2N$. 
%\begin{align}
%	&\Levciv[ce]\Levciv[ab]\int_x\int_y\int_u\int_v \Dl[](u,v) \K[ac](u,x) \K[eb](v,y)\dot{\Delta}_{l}(x,y)\\
%	=&\frac1{2}\int_u\int_v \Dl[uv]  \big(2i\chi^{\alpha ij}(u)\frac{\partial}{\partial u^{\alpha}} + i\cancel{\frac{\partial \chi^{\alpha ij}(u)}{\partial u^{\alpha}}} - \sqn\chi^{\alpha ik}(u)\chi_{\alpha}^{jk}(u) \big) \nonumber\\
%	&\big(2i\chi^{\beta ij}(v)\frac{\partial}{\partial v^{\beta}} + i\cancel{\frac{\partial \chi^{\beta ij}(v)}{\partial v^{\beta}}} - \sqn\chi^{\beta il}(v)\chi_{\beta}^{jl}(v) \big) \int_x\int_y\delta(u-x)\delta(v-y)\dot{\Delta}_{hxy}\\
%	=& \frac1{2}\int_x\int_y \Dl \big(2i\cmu\pmu - \sqn\cmu[ik]\cmd[jk] \big) \big(2i\cnu\pnu - \sqn\cnu[il]\cnd[jl] \big) \dot{\Delta}_{hxy}\\
%	=& -\frac1{2}\int_x\int_y \Dl \Big[ 4\cmu\cnu\pmu\pnu +\cancel{2i\sqn\cmu\cnu[ik]\cnd[jk]\pmu} +\cancel{2i\sqn\cnu\cmu[ik]\cmd[jk]\pnu} \nonumber\\
%	& -\frac1N\cmu[ik]\cmd[jk]\cnu[il]\cnd[jl]\Big]\dot{\Delta}_{hxy},
%\end{align}
\begin{align}
	&\Levciv[ce]\Levciv[ab]\int_x\int_y\int_u\int_v \Dl[](u,v) \K[ac](u,x) \K[eb](v,y)\dot{\Delta}_{l}(x,y)\nonumber\\
	&=\Levciv[ce]\Levciv[ab]\int_x\int_y\int_u\int_v \Dl[](u,v)\big[i\cmu(u) \{\levciv[dc]\gen[i][d][a]+\levciv[ad]\gen[i][d][c]\}\ddvar{u}{\mu} + \sqn\cmu[i](u)\cmd[j](u) \levciv[df]\gen[i][d][a] \gen[j][f][c]\big]\nonumber\\
	&\qquad \delta(u-x)\times \big[i\cnu[k](v) \{\levciv[gb]\gen[k][g][e]+\levciv[eg]\gen[k][g][b]\}\ddvar{v}{\nu} + \sqn\cnu[k](v)\cnd[l](v) \levciv[gh]\gen[k][g][e] \gen[l][h][b]\big]\nonumber\\
	&\qquad\delta(v-y)\dot{\Delta}_{l}(x,y)\nonumber\\
	&=\Levciv[ce]\Levciv[ab]\int_u\int_v \Dl[](u,v)\big[i\cmu(u) \{\levciv[dc]\gen[i][d][a]+\levciv[ad]\gen[i][d][c]\}\ddvar{u}{\mu} + \sqn\cmu[i](u)\cmd[j](u) \levciv[df]\gen[i][d][a] \gen[j][f][c]\big]\nonumber\\
	&\qquad \times \big[i\cnu[k](v) \{\levciv[gb]\gen[k][g][e]+\levciv[eg]\gen[k][g][b]\}\ddvar{v}{\nu} + \sqn\cnu[k](v)\cnd[l](v) \levciv[gh]\gen[k][g][e] \gen[l][h][b]\big]\dot{\Delta}_{l}(u,v)\nonumber\\
	&=\int_x\int_y\Dl[](x,y)\Big[4\cmu(x)\cnu[k](y)\Tr(T_iT_k)\ddvar{x}{\mu}\ddvar{y}{\nu}-2\frac{i}{\sqrt N}\Big(\cmu(x)\cnu[k](y)\cnd[l](y)\Tr(T_iT_kT_l)\ddvar{x}{\mu}\nonumber\\ &\qquad+\cmu(x)\cmd[j](x)\cnu[k](y)\Tr(T_iT_jT_k)\ddvar{y}{\nu}\Big) -\frac1N\cmu(x)\cmd[j](x)\cnu[k](y)\cnd[l](y)\Tr(T_iT_jT_kT_l)\Big]\dot{\Delta}_{l}(x,y),
	\label{ERGterm1}
\end{align}
where we've relabelled $u,v$ to $x,y$ in the third step. Recall that $T_i$ are the generators of the $j=1/2$ representation of $\mathfrak{su}(2)$. That is, $T_i=\frac12\sigma_i$ where $\sigma_i$ are the Pauli matrices. We have also made use above of the fact that for the generators, $\epsilon_{ab}T^b_c\eps^{cd}=(T^{-1})^d_a=T^d_a$, since Pauli matrices square to identity. The trace formulas for $T_i$ are
\begin{align}
	\Tr(T_iT_j)&=\frac12 \dij,\\
	\Tr(T_iT_jT_k)&=\frac i4\eps_{ijk},
	\label{3trace}\\
	\Tr(T_iT_jT_kT_l)&=\frac18(\dij\dij[kl]-\dij[ik]\dij[jl]+\dij[jk]\dij[il])
\end{align}
Thus, because the cubic terms in $\chi$ in \eqref{ERGterm1} are symmetric in two of the $\chi$s, on contracting with the $\eps_{ijk}$ resulting from the traces, they vanish.
Therefore, the \textbf{term 1} is
\begin{equation}
	\int_x\int_y\Dl[](x,y)\Big[2\cmu(x)\cnu(y)\ddvar{x}{\mu}\ddvar{y}{\nu} -\frac1{8N}\chi(x)^2\chi(y)^2\Big]
\end{equation}
The latter term in this is $O(1/N)$, so we do not look at it.
We write the leading term in momentum space,
\begin{equation}
	2\int_p\int_q\Dl[,p+q](q\cdot\chi^{ij}_p)(q\cdot\chi^{ij}_{-p})\dot{\Delta}_{h,q}
\end{equation}

\paragraph{Term 2} We look at the $O(1/\sqrt N)$ term from the $K^3$ term.
\begin{align}
	&\sqn \eps^{ab}\eps^{ce}\eps^{df} \int_x\int_y\int_z\int_u\int_v\int_w \dot{\Delta}_{l}(x,y)K_{ac}(u,x)\Dl[](u,w)K_{fe}(w,z)\Dl[](z,v)K_{db}(v,y)\nonumber\\
	&=\sqn \eps^{ab}\eps^{ce}\eps^{df} i^3\int_x\int_y\int_z\int_u\int_v\int_w\dot{\Delta}_{l}(x,y)\Dl[](u,w)\Dl[](z,v) (\levciv[gc]\gen[i][g][a]+\levciv[ag]\gen[i][g][c])\nonumber\\
	&(\levciv[hb]\gen[j][h][d]+\levciv[dh]\gen[j][h][b])
	(\levciv[oe]\gen[k][o][f]+\levciv[fo]\gen[k][o][e]) \cmu(u)\cnu[j](v)\chi^\rho_k(w)\nonumber\\
	&\ddvar{u}{\mu}\ddvar{v}{\nu}\ddvar{w}{\rho} \delta(u-x)\delta(v-y)\delta(w-z)\nonumber\\
	&=\frac{8i}{\sqrt N}\Tr(T_iT_jT_k)\int_x\int_y\int_z\cmu(x)\cnu[j](y)\chi^\rho_k(z) \Dl[](x,z)\ddvar{z}{\rho}\Dl[](z,y) \ddvar{x}{\mu}\ddvar{y}{\nu}\dot{\Delta}_{l}(x,y)\nonumber\\
	&=-\frac{2}{\sqrt N}\eps_{ijk}\int_x\int_y\int_z\cmu(x)\cnu[j](y)\chi^\rho_k(z) \Dl[](x,z)\ddvar{z}{\rho}\Dl[](z,y) \ddvar{x}{\mu}\ddvar{y}{\nu}\dot{\Delta}_{l}(x,y),
\end{align}
where we've made use of the trace formula \eqref{3trace}.
%In the following, $\partial\cdot\chi$ terms in $\K$ are dropped. Terms with index $\mu$ are associated with $x$, $\nu$ with $y$, $\rho$ with $z$.
%\begin{align}
%	& -\frac12\frac{1}{\sqrt N}\int_x\int_y \dot{\Delta}_{hxy} \int_u\int_v \K[ik](u,x)\K[jl](v,y)\int_z\int_w\Dl[uw] \K[kl](w,z)\Dl[zy]\nonumber\\
%	= & -\frac12\frac{1}{\sqrt N} \int_u\int_v \int_w \Dl[uw]\big(2i\chi^{\alpha ij}(u)\frac{\partial}{\partial u^{\alpha}} - \sqn\chi^{\alpha ik}(u)\chi_{\alpha}^{jk}(u) \big) \nonumber\\
%	&\big(2i\chi^{\beta ij}(v)\frac{\partial}{\partial v^{\beta}}- \sqn\chi^{\beta il}(v)\chi_{\beta}^{jl}(v) \big) \Big(\int_x\int_y\delta(u-x)\delta(v-y)\dot{\Delta}_{hxy}\Big)\nonumber\\ 
%	&\big(2i\chi^{\gamma ij}(w)\frac{\partial}{\partial w^{\gamma}}- \sqn\chi^{\gamma in}(w)\chi_{\gamma}^{jn}(w) \big) \Big(\int_z \delta(w-z) \Dl[zy]\Big)\\
%	=&-\frac12\frac{1}{\sqrt N}\int_x\int_y\int_z \Dl[xz](2i\cru[kj]\pro-\sqn\cru[kl]\crd[jl])\Dl[zy] (2i\cmu[ik]\pmu-\sqn\cmu[im]\cmd[km])\nonumber\\ &(2i\cnu[ij]\pnu-\sqn\cnu[in]\cnd[jn])\dot{\Delta}_{hxy}.
%\end{align}

In momentum space, the $O(1/\sqrt N)$ term is
%\begin{equation}
%	-\frac{4}{\sqrt N}\int_p\int_q\int_r\Dl[,r+p]\Dl[,r-q]\dot{\Delta}_{l,r} (q-r)\cdot\chi^{kj}_{-p-q}r\cdot\chi^{ik}_pr\cdot\chi^{ij}_q,
%\end{equation}
% \begin{align}
	% 	&-\frac12\frac{1}{\sqrt N}\int_p\int_q\int_r\Dl[,r]\Dl[,r-q]\dot{\Delta}_{h,r+p} \Big[-8(r-q).\chi^{kj}_q(r+p).\chi^{ik}_p(r+p).\chi^{ij}_{-p-q} \nonumber\\
	% 	&-4\sqn(r-q).\chi^{kj}_q(r+p).\chi^{ik}_p\cnu[in]\cnd[jn] +4\sqn(r+p).\chi^{ik}_p(r+p).\chi^{ij}_{-p-q}\cru[kl]\crd[jl] \nonumber\\
	% 	&+4\sqn(r-q).\chi^{kj}_q(r+p).\chi^{ij}_{-p-q}\cmu[im]\cmd[km] \nonumber\\ 
	% 	&+\frac2N(r-q).\chi^{kj}_q\cmu[im]\cmd[km]\cnu[in]\cnd[jn] +\frac2N(r+p).\chi^{ik}_p\cru[kl]\crd[jl]\cnu[in]\cnd[jn] \nonumber\\
	% 	&-\frac2N(r+p).\chi^{ij}_{-p-q}\cmu[im]\cmd[km]\cru[kl]\crd[jl] \nonumber\\
	% 	&-\frac{1}{N\sqrt N}\cmu[im]\cmd[km]\cru[kl]\crd[jl]\cnu[in]\cnd[jn]\Big],
	% \end{align}
%where the index $\mu$ corresponds to $\chi_p$, $\rho$ to $\chi_q$ and $\nu$ to $\chi_{-p-q}$. This can be modified a bit by using transversality.
\begin{equation}
	=\frac{2i}{\sqrt N}\eps_{ijk}\int_p\int_q\int_r\dpqr\int_k\Dl[,k+p]\Dl[,k-q]\dot{\Delta}_{l,k} (k.\chi_k(q))(k.\chi_i(p))(k+p).\chi_j({r}),
\end{equation}
%till here all ok 7/6/22
% \begin{align}
	% 	=&-\frac12\frac{1}{\sqrt N}\int_p\int_q\int_r\Dl[,r]\Dl[,r-q]\dot{\Delta}_{h,r+p} \Big[8(r.\chi^{kj}_q)(r.\chi^{ik}_p)(r+p).\chi^{ij}_{-p-q} \nonumber\\
	% 	&+4\sqn(r.\chi^{kj}_q)(r.\chi^{ik}_p)\cnu[in]\cnd[jn] +4\sqn(r.\chi^{ik}_p)(r+p).\chi^{ij}_{-p-q}\cru[kl]\crd[jl] -4\sqn(r.\chi^{kj}_q)(r+p).\chi^{ij}_{-p-q}\cmu[im]\cmd[km] \nonumber\\ &-\frac2N(r.\chi^{kj}_q)\cmu[im]\cmd[km]\cnu[in]\cnd[jn] +\frac2N(r.\chi^{ik}_p)\cru[kl]\crd[jl]\cnu[in]\cnd[jn] -\frac2N(r+p).\chi^{ij}_{-p-q}\cmu[im]\cmd[km]\cru[kl]\crd[jl] \nonumber\\
	% 	&-\frac{1}{N\sqrt N}\cmu[im]\cmd[km]\cru[kl]\crd[jl]\cnu[in]\cnd[jn]\Big],
	% 	\label{order3}
	% \end{align}
%Of these, we keep only the $O(1/\sqrt{N})$ terms.

As can be seen from \eqref{pathintegral}, we can replace $\cmd(x)$ with $-i\delta/\delta\smu(x)$, (or $\cmd(p)$ with $-i\delta/\delta\smu(-p)$), so that the flipped ERG equation \eqref{ergphi} looks like this in terms of $\smu$, after setting $\phi_h$ to $0$.
\begin{equation}
	\boxed{
		\begin{aligned}
			\ddt\e{-S_{\lm}[\smu]}  =& \bigg\{-2\int_x\int_y \Dl[](x,y) \frac{\partial \dot{\Delta}_{l}(x,y)}{\partial x^\mu \partial y^\nu}\frac{\delta^2}{\delta \smd(x) \delta\snd(y)}
			%	\nonumber
			\\
			&-\frac{2i}{\sqrt N}\eps_{ijk}\int_x\int_y\int_z \Dl[](x,z)\frac{\partial \Dl[](y,z)}{\partial z^\rho} \frac{\partial \dot{\Delta}_{l}(x,y)}{\partial x^\mu \partial y^\nu}\frac{\delta^3}{\delta \smd[i](x)\delta\snd[j](y)\delta\srd[k](z)}\bigg\} \e{-S_{\lm}[\smu]} +\ldots
			\label{ergsigma}\\
			=& \bigg\{-2\int_p\int_q\Dl[,p+q] \dot{\Delta}_{l,q}q^{\mu}q^{\nu}\frac{\delta^2}{\delta \smu(p) \delta\snu(-p)}
			%	\nonumber
			\\
			& +\frac{2}{\sqrt N}\eps_{ijk}\int_p\int_q\int_r\dpqr\int_k\Dl[,k+p]\Dl[,k-q]\dot{\Delta}_{l,k} k_{\mu}k_{\nu}(k+p)_{\rho}\\
			&\frac{\delta^3}{\delta \smd(-p)\delta\snd[j](-q)\delta\srd(-r)}\bigg\}
			\e{-S_{\lm}[\smu]} +\ldots,
			%	\label{ergsigmamom}
		\end{aligned}
	}
\end{equation}
where we've written down the leading and subleading order terms in $1/\sqrt N$.

The solution of this equation, $S_\Lambda[\sigma]$, is an action for $\sigma$ which gives approximate results for momenta below $\Lambda$. When $\Lambda\to\Lambda_b$, it becomes exact, i.e., the correlations of $\sigma$ calculated with it are exact. 

We need the form of $S_\lm[\smu]$ for the functional derivative to act on. Remember that $\phi_h$ has been set to zero now. We have
\begin{equation}
	\e{-S_{\Lambda}[\smu[]]} =  \int\CD\chi \exp\Big\{i\int_x\cmu\smd \Big\}\exp\{-\sil[0,\cmd[]]\},
	\label{sigaction}
\end{equation}
where
\begin{align}
	\sil[0,\cmd[]]=&	\frac 12\Tr\log\Big[\skewsym(\sqrt{N}\dli\epsilon_{ab} - \K)\Big]\nonumber\\
	=&	\frac 12\Tr\Big[\log(\sqrt{N}\skewsym\dli\epsilon_{ab})-\sqn\delta^{IJ}\Delta_l\epsilon^{ab} \K[bc]-\frac1{2N}\delta^{IJ}\Delta_l\eps^{ab} \K[bc]\Delta_l \eps^{cd}\K[de]-\ldots\Big]\nonumber\\
	=&-\sqrt N\int_x\int_y \Dl[](x,y)\Levciv[ab]\K[ba](y,x)\nonumber\\ &-\frac12\int_x\int_y\int_z\int_w \Dl[](x,y)\eps^{ab}\K[bc](y,z)\Dl[](z,w)\eps^{cd}\K[da](w,x)+\ldots\nonumber\\
	=& -\frac{1}{2}\int_x \Dl[](0) \cmu\cmd +\int_x\int_y\ddvar{x}{\mu}\Dl[](x,y)\ddvar{y}{\nu}\Dl[](y,x)\cmu(x)\cnu(y)+\ldots\nonumber\\
	\equiv& \frac{1}{2}\int_x\int_y\cmu(x)\I(x,y)\cnu(y)+\ldots,
	\label{WilsonActionKinetic}
\end{align}
where we've kept only the kinetic term, (higher order terms in $\cmu$ are also higher order in $1/\sqrt N$), and  
\begin{equation}
	\I(x,y) = -\Delta_l(0)\dxy\dmn + 2\pmu\Dl[](x,y)\pnu\Dl[](y,x).
\end{equation}
Here, $I$ is the \textit{low energy} propagator for the $\chi$ field, because we get it after integrating out low energy modes $\fil$.
In momentum space, the integral looks like 
\begin{equation}
	\sil[0,\cmd[]] = \frac1{2}\int_p\int_q\cmu(p)\I(p,q)\cnu(q)+\ldots,
\end{equation}
with
\begin{align}
	\I(p,q) =& -\dmn\delta(p+q)\int_r \Dl[,r] +2\delta(p+q)\int_rr_{\mu}(p-r)_{\nu}\Dl[,r]\Dl[,p+r]\nonumber\\
	=& -\dmn\delta(p+q)\int_r \Dl[,r] -2\delta(p+q)\int_rr_{\mu}r_{\nu}\Dl[,r]\Dl[,p+r]
	\label{Imom}\\
	=& -\dmn\delta(p+q)\Big\{\int_r \Dl[,r] +\frac2D\int_rr^2\Dl[,r]\Dl[,p+r]\Big\}\nonumber\\
	\equiv&\dmn\delta(p+q)I(p^2).
	\label{Isigmapropagator}
\end{align}
Here $\Dl[,r]$ is the Fourier transform of the $\Dl[]$ we had earlier, and $D$ is the number of dimensions of the boundary manifold. We've also made use of transversality to get rid of $p$ in the expression. 
Completing the gaussian from $\eqref{sigaction}$, 
\begin{equation}
	S_{\Lambda}[\smu[]]=\frac{1}{2}\int_p\frac{1}{I(p^2)}\smu(p)\smd(-p)=\frac{1}{2}\int_x\int_yI^{-1}_{\mu\nu}(x,y)\smu(x)\snu(y).
	\label{o1waction}
\end{equation}
%While for $\fih$, the quantity $S_\Lambda[\fih]$ has no interpretation as a Wilsonian action, $S_\Lambda[\smu]$ is a legitimate action for the $\sigma$ modes of momenta below cutoff $\Lambda$.
We have to check if this obeys the ERG equation \eqref{ergsigma} to leading order.
\begin{equation}
	LHS=\ddt\e{-S_{\lm}[\smu]} =\frac{1}{2}\int_p\frac{\dot{I}(p^2)}{I(p^2)^2}\smu(p)\smd(-p)\e{-S_{\lm}[\smu]}.
\end{equation}
\begin{align}
	RHS=& -2\int_p\int_q\Dl[,p+q] \dot{\Delta}_{l,q}q^{\mu}q^{\nu}\frac{\delta^2}{\delta \smu(p) \delta\snu(-p)} \e{-S_{\lm}[\smu]}\\
	=& -2\int_p\int_q\Dl[,p+q] \dot{\Delta}_{l,q}q^{\mu}q^{\nu}\times 2\times\frac12 \frac{1}{I(p^2)}\smd(-p)\times 2\times\frac12 \frac{1}{I(p^2)}\snd(p)\\
	=& \frac{1}{2}\int_p\frac{\dot{I}(p^2)}{I(p^2)^2}\smu(p)\smd(-p)\e{-S_{\lm}[\smu]} =LHS,
\end{align}
since, from \eqref{Imom}, 
\begin{align}
	\Dot{I}(p^2)= &-2\int_rr_{\mu}r_{\nu}\Dl[,r]\dot{\Delta}_{l,p+r}-2\int_rr_{\mu}r_{\nu}\Dl[,p+r]\dot{\Delta}_{l,r}\\
	=&-2\int_r(r-p)_{\mu}(r-p)_{\nu}\Dl[,r-p]\dot{\Delta}_{l,r}-2\int_rr_{\mu}r_{\nu}\Dl[,p+r]\dot{\Delta}_{l,r}\\
	=&-2\int_r(r+p)_{\mu}(r+p)_{\nu}\Dl[,r+p]\dot{\Delta}_{l,r}-2\int_rr_{\mu}r_{\nu}\Dl[,p+r]\dot{\Delta}_{l,r},
\end{align}
where we've used the fact that $\Dl[,r]=\Dl[,-r]$. The $p_\mu,p_\nu$ in this drop away on contracting with $\sigma$, and we get $4\int_rr_{\mu}r_{\nu}\Dl[,p+r]\dot{\Delta}_{l,r}$

%\paragraph{ignore below}
%\begin{align}
%	RHS =& 2\int_x\int_y \Dl \pmu\pnu\dot{\Delta}_{hxy}\frac{\delta^2}{\delta \smd \delta\snd}\e{-S_{\lm}[\smu]}\\
%	=&  -2\int_x\int_y \Dl \pmu\pnu\dot{\Delta}_{hxy}\frac{\delta}{\delta \smd}\bigg(\int_zI^{-1}_{\rho\nu}(z,y)\sru(z)\e{-S_{\lm}[\smu]}\bigg)\\
%	=& 2\int_x\int_y \Dl \pmu\pnu\dot{\Delta}_{hxy}\Big(\int_zI^{-1\nu}_{\rho}(z,y)\sru(z)\Big)\Big(\int_wI^{-1\mu}_{\kappa}(x,w)\sigma^{\kappa ij}(w)\Big)\e{-S_{\lm}[\smu]}\\
%	=& -2\int_x\int_y \pnu\Dl \pmu\dot{\Delta}_{hxy}\Big(\int_zI^{-1\nu}_{\rho}(z,y)\sru(z)\Big)\Big(\int_wI^{-1\mu}_{\kappa}(x,w)\sigma^{\kappa ij}(w)\Big)\e{-S_{lm}[\smu]}\\
%	=& 2\int_x\int_y \pnu\Dl[yx] \pmu\dot{\Delta}_{hxy}\Big(\int_zI^{-1\nu}_{\rho}(z,y)\sru(z)\Big)\Big(\int_wI^{-1\mu}_{\kappa}(x,w)\sigma^{\kappa ij}(w)\Big)\e{-S_{\lm}[\smu]},
%\end{align}
%where we've transferred the derivative, and used the fact that $\int_zI^{-1\nu}_{\rho}(z,y)\sru(z)$ is transverse. In the last step, we've used the fact that $\pnu\Dl[yx]=-\pnu\Dl$, as can be seen from writing $\Dl=\int_p \e{i(x-y).p}\Dl[,p]$
%Since
%\begin{equation}
%	\dot{I}_{\mu\nu}(x,y)=2\pmu\dot{\Delta}_{hxy}\pnu\Dl[yx] +2\pmu\Dl\pnu\dot{\Delta}_{hyx},
%\end{equation}
%\begin{equation}
%	RHS=-\frac12\int_x\int_y\smu\Big(\int_z\int_wI^{-1}_{\mu\rho}(x,z)\dot I^{\rho\kappa}(z,w)I^{-1}_{\kappa\nu}(w,y)\Big)\snu\e{-S_{\lm}[\smu]}=LHS.
%\end{equation}
%
%\paragraph{resume consideration}\hfill

For the next order term in the flipped ERG equation, we have from \eqref{ergsigma},
\begin{align}
	&\frac{2}{\sqrt N}\eps_{ijk}\int_p\int_q\int_r\dpqr\int_k\Dl[,k+p]\Dl[,k-q]\dot{\Delta}_{l,k} k_{\mu}k_{\nu}(k+p)_{\rho}\frac{\delta^3}{\delta \smd(-p)\delta\snd[j](-q)\delta\srd(-r)}\nonumber\\
	& \exp\Big\{-\frac{1}{2}\int_s\frac{1}{I(s^2)}\sigma^{\kappa }_{i^\prime}(s)\sigma_{\kappa}^{i^\prime}(-s)\Big\}\nonumber\\
	=& \frac{2}{\sqrt N}\eps_{ijk}\int_p\int_q\int_r\dpqr\int_k\frac{\Dl[,k+p]\Dl[,k-q]\dot{\Delta}_{l,k}}{I(p^2)I(q^2)I(r^2)} k_{\mu}k_{\nu}(k+p)_{\rho}\smu(p)\snu[j](q)\sru(r)\nonumber\\
	& \exp\Big\{-\frac{1}{2}\int_s\frac{1}{I(s^2)}\sigma^{\kappa }_{i^\prime}(s)\sigma_{\kappa}^{i^\prime}(-s)\Big\}.
	\label{cubicsigma}
\end{align}
The flipped ERG equation for $\sigma$ is
\begin{equation}
	\boxed{
		\begin{aligned}
			\ddt\e{-S_{\lm}[\smu]}  
			=& \bigg\{\frac12\int_p \dot{I}(p^2)\frac{\delta^2}{\delta \smu(p) \delta\smd(-p)}\\
			& +\frac{2}{\sqrt N}\eps_{ijk}\int_p\int_q\int_r \dpqr\int_k\frac{\Dl[,k+p]\Dl[,k-q]\dot{\Delta}_{l,k}}{I(p^2)I(q^2)I(r^2)} k_{\mu}k_{\nu}(k+p)_{\rho}\smu(p)\snu[j](q)\sru(r) \bigg\}\\
			& \e{-S_{\lm}[\smu]} +\ldots
	\end{aligned}}
	\label{ERG}
\end{equation}

\section{Solution of the flipped ERG equation}
\label{solutionFlippedERG}
The flipped ERG equation doesn’t have the physical interpretation of Wilsonian RG evolution.
It is closer in spirit to ordinary perturbative field theory calculations where a UV regulator $\lm$ is
introduced and all momenta below $\lm$ are integrated and eventually the limit $\lm \to \infty$ is taken (with
possible addition of counterterms if there are divergences). It is also closer in spirit to AdS/CFT
calculations where the bulk fields are integrated from $z = \infty$ to $z = \eps $ and subsequently the $\eps \to 0$
limit is taken.

In the present case the $\phi$ fields below $\lm$ are being integrated, keeping $\sigma^\mu$ fixed. When $\lm \to \infty$
one obtains an action for $\sigma^\mu$ 
 that can be used for calculating  correlations of $\sigma^\mu$. For finite $\lm$ only momentum modes of $\phi$ below $\lm$ have been integrated. Consequently only low momentum correlations of $\sigma^\mu$ fields can be calculated. Note that this cannot be interpreted as a conventional Wilson action for $\sigma^\mu$. Since it has no information about physics above $\lm$. 
It is as if there is a physical cutoff at $\lm$. As $\lm \to \infty$ the full $\sigma$ action is obtained. Note that this is to be interpreted as a bare action for $\sigma^\mu$ because it contains all momentum modes - nothing has been integrated. If one now integrates out the $\sigma^\mu$ modes with a source one can obtain the Generating Functional $W[J^\mu]$ where $J^\mu$ is a source for $\sigma ^\mu$. This can be obtained by solving the bulk equations by adding a source for $\sigma^\mu$ at the boundary rather than fixing the value of $\sigma^\mu$ at the boundary.

Picking up from \eqref{WilsonActionKinetic}, we can write down the flipped ERG action to next order in $1/N$. 
\begin{align}
\sil[0,\cmd[]]=&	\frac 12\Tr\log\Big[\skewsym(\sqrt{N}\dli\epsilon_{ab} - \K)\Big]\nonumber\\
=&	\frac 12\Tr\Big[\log(\sqrt{N}\skewsym\dli\epsilon_{ab})-\sqn\delta^{IJ}\Delta_l\epsilon^{ab} \K[bc]-\frac1{2N}\delta^{IJ}\Delta_l\eps^{ab} \K[bc]\Delta_l \eps^{cd}\K[de]\nonumber\\
&-\frac1{3N\sqrt N}\delta^{IJ}\Delta_l\eps^{ab} \K[bc]\Delta_l \eps^{cd}\K[de]\Dl\Levciv[ef]\K[fg]-\ldots\Big]
\end{align}
The cubic term is 
\begin{equation}
-\frac{1}{3\sqrt{N}}\int_{1-6}\Dl[12]\Levciv\K[bc][23]\Dl[34]\Levciv[cd]\K[de][45]\Dl[56]\Levciv[ef]\K[fa][61],
\end{equation}
where we're representing coordinates $x_1,x_2,x_3,x_4,x_5,x_6$ by their subscripts. From the definition of $\K$ in \eqref{Kab}, keeping only the $O(\sqn)$ term, 
\begin{align}
=&-\frac{1}{3\sqrt{N}} \int_{1-6}\Dl[12] 2i\chi^{\mu}_i(x_2)\gen\partial_{2\mu}\delta_{23} \Dl[34] 2i\chi^{\nu}_j(x_4)\gen[j][c][e]\partial_{4\nu}\delta_{45}\Dl[56] 2i\chi^{\rho}_k(x_6)\gen[k][e][a]\partial_{6\rho}\delta_{61}\nonumber\\
=&\frac{8i}{3\sqrt{N}} \Tr(T_iT_jT_k) \int_{x}\int_y\int_z\chi^{\mu}_i(x)\chi^{\nu}_j(y) \chi^{\rho }_k(z) \pmu\Dl(x,y)\pnu\Dl(y,z)\ddvar{z}{\rho}\Dl(z,x) \nonumber\\
=&-\frac{2}{3\sqrt{N}} \eps_{ijk} \int_{x}\int_y\int_z\chi^{\mu}_i(x)\chi^{\nu}_j(y) \chi^{\rho }_k(z) \pmu\Dl(x,y)\pnu\Dl(y,z)\ddvar{z}{\rho}\Dl(z,x) \nonumber\\
=& \frac{2i}{3\sqrt{N}} \eps_{ijk}\int_p\int_q\int_r\dpqr\int_k\Dl[,k+p]\Dl[,k-q]{\Delta}_{h,k} (k.\chi^{i}_p)(k.\chi^{j}_q)(k+p).\chi^{k}_{r}\nonumber\\
=& \frac{2}{3\sqrt{N}} \eps_{ijk}\int_p\int_q\int_r\dpqr\int_k\Dl[,k+p]\Dl[,k-q]{\Delta}_{h,k} k_{\mu}k_{\nu}(k+p)_{\rho}\frac{\delta^3}{\delta \smd(-p)\delta\snd[j](-q)\delta\srd(-r)}.
%	\\
%	=& \frac{4}{3\sqrt{N}}\int_{246}\frac{\delta^3}{\delta\sigma^{ij}_\alpha(x_2)\sigma^{ jk}_\beta(x_4)\sigma^{ ki}_\gamma(x_6)} \partial_\alpha\Dl[24]\partial_\beta\Dl[46]\partial_\gamma\Dl[62]\\
%	=& \frac{4}{3\sqrt{N}}\int_{246}\frac{\delta^3}{\delta\sigma^{ij}_\alpha(x_2)\sigma^{ jk}_\beta(x_4)\sigma^{ ki}_\gamma(x_6)} \partial_\alpha\Dl[24]\partial_\beta\Dl[46]\partial_\gamma\Dl[62]
\end{align}
In the above, we have made use of the trace formula \eqref{3trace}, and substituted for $\cmd(p)$ $-i\delta/\delta\smu(-p)$ which follows from \eqref{pathintegral}.

The flipped ERG action is then given from \eqref{sigaction} by
\begin{equation}
\e{-S_\Lambda[\sigma^\mu]} = \e{-\textrm{cubic term in }\delta/\delta\sigma^\mu}\e{-\textrm{kinetic term \eqref{o1waction}}},
\end{equation} 
so the cubic term in $S_\Lambda$ is
\begin{align}
-\frac{2}{3\sqrt N}\eps_{ijk} \int_p\int_q\int_r\dpqr\int_k \frac{\Dl[,k+p]\Dl[,k-q]{\Delta}_{h,k}}{I(p^2)I(q^2)I(r^2)}  k_{\mu}k_{\nu}(k+p)_{\rho}\smu(p)\snu[j](q)\sru(r).
\end{align}
%The $O(\sqn)$ term is
%\begin{equation*}
%	\frac{4i}{3\sqrt N}\bigg\{\ddt\int_p\int_q\int_r\frac{\Dl[,r+p]\Dl[,r-q]\Dl[,r]}{I(p^2)I(q^2)I((p+q)^2)} r_{\mu}r_{\nu}(r+p)_{\rho}\smu[ik](p)\snu[kj](q)\sru(-p-q)\bigg\} \e{-S_{\lm}[\smu]}.
%\end{equation*}
%Since the L.H.S. is $-\ddt (S_{\lm}[\smu])\e{-S_{\lm}[\smu]}$,
The flipped ERG action is therefore
\begin{equation}
\boxed{
	\begin{aligned}
	S_{\Lambda}[\smu[]]=&\frac{1}{2}\int_p\frac{1}{I(p^2)}\smu(p)\smd(-p)\\
	&-\frac{2}{3\sqrt N}\eps_{ijk}\int_p\int_q\int_r\dpqr\int_k \frac{\Dl[,k+p]\Dl[,k-q]{\Delta}_{h,k}}{I(p^2)I(q^2)I(r^2)}  k_{\mu}k_{\nu}(k+p)_{\rho}\smu(p)\snu[j](q)\sru(r),
	\end{aligned}
}
\label{wilsonaction}
\end{equation}
and it satisfies the flipped ERG equation \eqref{ERG}.

%%%%%%%%%%%%%%%%%%%%%%%%%%%%%%%%%%%%%%%%%%%%%%%%%%%%%%%%%%%%%%%%%%%%%%%%%%%%%%%
\section{The boundary action from the bulk action}
To verify the nonlocal holographic action given in \eqref{bulkaction}, we solve it classically, and substitute the classical solution for the field in the action to obtain the boundary action.
\paragraph{To the leading order in $\boldsymbol{N}$}
The equations of motion are
\begin{equation}
\ddt \frac{\dsmu}{\dot{I}}=0 \implies  \frac{\dsmu(p)}{\dot{I}(p^2)}\equiv b^{\mu }_i(p),
\label{KEOM}
\end{equation}
for some constant $b^{\mu}_i$.
Thus, 
\begin{equation}
b^{\mu}_i(p)=\frac{\smu(t,p)-\sigma_{bi}^{\mu}(p)}{I(t,p^2)-I_b(p^2)},
\label{constb}
\end{equation}
for any $t$. The subscript $b$ stands for \textit{bare}, and indicates quantitites at the boundary.

As an aside, vanishing of the variation of action also implies the following condition.
\begin{align}
&\int_{t_0}^{\bar{t}} \der t\int_p \ddt\bigg(\delta \smu(t,p) \frac{\dsmd(t,p)}{\dot{I}(t,p^2)}\bigg)+\int_p \delta \sigma^{\mu}_{bi}(p) \frac{\sigma^{i}_{b\mu}(p)}{I_b(p^2)}=0\\
\implies &\int_{t_0}^{\bar{t}} \der t\int_p \ddt(\delta \smu(t,p) b_\mu^{i}(p))+\int_p \delta \sigma^{\mu}_{bi}(p) \frac{\sigma^{i}_{b\mu}(p)}{I_b(p^2)}=0 \nonumber\\
\implies &  -\int_p \delta \sigma^{\mu}_{bi}(p)b_\mu^{i}(p) + \int_p \delta \sigma^{\mu }_{bi}(p) \frac{\sigma^{i}_{b\mu}(p)}{I_b(p^2)} =0 \nonumber\\
\implies & b^{\mu }_i(p) =  \frac{\sigma^{\mu }_{bi}(p)}{I_b(p^2)} .
\label{constantRatio}
\end{align}
Then we have from \eqref{constb}
\begin{equation}
\smu(p,z)=b^\mu_i(p)I(p,z).
\label{EOM}
\end{equation}

The classical bulk action can then be written, (in terms of $\smu(t_f)\equiv \sigma_{fi}^{\mu}$ and $\smu(t_0)\equiv\sigma_{bi}^{\mu}$), as
\begin{align}
S_{cl}^{(0)}[\sigma^\mu_f,\sigma_b^\nu]= &\frac12\int_{t_0}^{t_f}\der t\int_p \frac{\dsmu(p)}{\dot{I}(p^2)}\dsmd(-p)=\frac12\int_{t_0}^{t_f}\der t\int_p b^{\mu }_i\dsmd(-p) \nonumber \\
=&  \frac12\int_p b_\mu^{ i}(p)\big(\sigma_{fi}^{\mu }(-p) - \sbu(-p)\big)\nonumber\\
=&  \frac12\int_p \frac{\big(\sfu(p)-\sigma_{bi}^{\mu}(p)\big)^2}{I_f(p^2)-I_b(p^2)},
\end{align}
from \eqref{constb}. $I_f(p^2)\equiv I(t_f,p^2)$ and $I_b(p^2)=I(t_0,p^2)$. Since $\Dl$ vanishes for $t=t_0$, $I_b=0$. We also choose to set $\sbu=0$, since it must vanish in the limit $\Lambda_0\to\infty$, and we do not gain anything by choosing a general $\sbu$ for a finite $\Lambda_0$. Thus,
\begin{equation}
S_{cl}^{(0)}[\sigma^\mu_f]=\frac12\int_p \frac{\sfu(p)\sfd(-p)}{I_f(p^2)},
\end{equation}
which matches with the Wilson action in \eqref{wilsonaction} to leading order.

\paragraph{$\boldsymbol{O(1/\sqrt N)}$}
Again we look at classical solution, including next order terms in EOM. First we write the the solution as
\begin{equation}
\sigma^\mu\equiv \sigma^\mu_0 + \sqn \sigma_1^\mu + O\Big(\frac1N\Big).
\end{equation}
The equation of motion at order $1/\sqrt N$ is (see \eqref{bulkaction})
\begin{equation}
 \ddt \frac{\dot{\sigma}_{1\mu}^{i}(-p)}{\dot{I}(p^2)} = 6\eps_{ijk}\int_q\int_r\dpqr\mathcal{V}_{\mu\nu\rho}(p,q,r;t)\snu[0j](q)\sru[0k](r),
\label{eomorder1}
\end{equation}
where we've decluttered the equations by introducing
\begin{equation}
\mathcal{V}_{\mu\nu\rho}(p,q,r;t)=\int_k \frac{\Dl[,k+p]\Dl[,k-q]\dot{\Delta}_{l,k}}{I(p^2)I(q^2)I(r^2)} k_{\mu}k_{\nu}(k+p)_{\rho}.
\end{equation}
We'll also use 
\begin{equation}
\mathcal{U}_{\mu\nu\rho}(p,q,r;t)=\int_k \frac{\Dl[,k+p]\Dl[,k-q]{\Delta}_{l,k}}{I(p^2)I(q^2)I(r^2)} k_{\mu}k_{\nu}(k+p)_{\rho}.
\end{equation}
Note that 
\begin{align}
\ddt\int_p\int_q\int_r\dpqr \mathcal{U}_{\mu\nu\rho}&(p,q,r;t)\sigma^\mu(p)\sigma^\nu(p)\sigma^\rho(r)\nonumber\\
&=3\int_p\int_q\int_r\dpqr \mathcal{V}_{\mu\nu\rho}(p,q,r;t)\sigma^\mu(p)\sigma^\nu(p)\sigma^\rho(r),
\end{align}
when $\sigma^\mu$ is transverse.

We can use this relation to substitute for the order 1 field in terms of order 0 field in the action like so. The order 1 part of the kinetic term is
\begin{equation}
\int_{t_0}^{t_f}\der t\int_p \frac{\dsmu[1i](-p)}{\dot{I}(p^2)}\dot{\sigma}_{0\mu}^{i}(p) =-\int_{t_0}^{t_f}\der t\int_p \ddt\bigg(\frac{\dsmu[0i](-p)}{\dot{I}(p^2)}\bigg)\sigma_{1\mu}^{i}(p) 
+\bigg[\int_p\frac{\dsmu[0i](-p)}{\dot{I}(p^2)}\sigma_{1\mu}^{i}(p)\bigg]_{t_0}^{t_f}
\end{equation}
We choose the boundary conditions $\sigma_1^\mu=0$ at both $t=t_0$ and $t=t_f$. Therefore the surface term vanishes. The first term also vanishes because $\frac{\dsmu[i0](-p)}{\dot{I}(p^2)}=b^\mu$ is a constant.

Thus the only term at this order in the EOM is the cubic term in the action at $O(1/\sqrt N)$ (see \eqref{bulkaction}), given by
\begin{align*}
-2\eps_{ijk}\int_{t_0}^{t_f}\der t\int_p\int_q\mathcal{V}_{\mu\nu\rho}(p,q;t)&\sigma_{0i}^{\mu }(p)\snu[0j](q)\sru[0k](-p-q)\\
&=-\frac{2}3\eps_{ijk} \int_p\int_q\mathcal{U}_{\mu\nu\rho}(p,q;t_f)\sigma_{f0}^{\mu i}(p)\snu[f0j](q)\sru[f0k](-p-q).
\end{align*}
Thus, the classical bulk action is 
\begin{equation}
\boxed{
	S_{cl}[\sigma^\mu_{f0}]=\frac12\int_p \frac{\sigma_{f0i}^{\mu}(p)\sigma_{f0\mu}^{ i}(p)}{I_f(p^2)} - \frac{2}{3\sqrt N}\eps_{ijk} \int_p\int_q\mathcal{U}_{\mu\nu\rho}(p,q;t_f)\sigma_{f0i}^{\mu }(p)\snu[f0j](q)\sru[f0k](-p-q).}
\end{equation}
This matches exactly with the Wilson action \eqref{wilsonaction}. 
The classical action is only in terms of $\sigma^\mu_{f0}$ because it is the only free parameter---all the excitations in $\sigma_f$ are completely determined in terms of $\sigma^\mu_{f0}$ because of the way we've chosen our boundary conditions. 

%%%%%%%%%%%%%%%%%%%%%%%%%%%%%%%%%%%%%%%%%%%%%%%%%%%%%%%%%%%%%%%%%%%%%%%%%%%%%%%

\section{Cubic term $z$-dependent part}
\label{integral}
The integral to be calculated is 
\begin{equation}
	\mathcal{K}_{\mu\nu\rho}(p,q,r) = \int \frac{\mathrm dk^D}{(2\pi)^D}\frac{1}{k^{2a_1}}\frac{1}{(k+p)^{2a_2}}\frac{1}{(k-q)^{2a_3}}k_\mu k_\nu (k+p)_\rho.
\end{equation}
The regularization scheme we leave unspecified for now. We are eventually interested in $a_i=1$ and $D=3$.
\begin{equation}
	\mathcal{K}_{\mu\nu\rho}(p,q,r)=\int ds_1\int ds_2\int ds_3 \frac{s_1^{a_1-1}s_2^{a_2-1}s_3^{a_3-1}}{\Gamma(a_1)\Gamma(a_2)\Gamma(a_3)}\int \dk e^{-(k-q)^2s_3}e^{-(k+p)^2s_2}e^{-k^2s_1}k_\mu k_\nu (k+p)_\rho
\end{equation}
%At this point let us introduce the UV regulator used in ERG of the form $e^{-bp^2}$ for each propagator, for some appropriate $b$. Then the integral becomes, 
Setting $a_i=1$,
\begin{align*}
	&\mathcal{K}_{\mu\nu\rho}(p,q,r)\\
	=&\int ds_1\int ds_2\int ds_3 \int \dk \exp\{-(k-q)^2s_3-(k+p)^2s_2-k^2s_1\}k_\mu k_\nu (k+p)_\rho.
%	&r\to k\equiv r+xp-yq;\quad \ell\equiv s_1+s_2+s_3+3b;w\equiv s_1+b;x\equiv s_2+b;y\equiv s_3+b\\
\end{align*} 
We make a transformation of variables to get a known integral. 
\begin{align}
	& \ell\equiv s_1+s_2+s_3;\beta_1\equiv \frac{s_2s_3}{\ell};\beta_2\equiv \frac{s_3s_1}{\ell};\beta_3\equiv \frac{s_1s_2}{\ell}\nonumber\\
	&k\to k'\equiv k+\ell^{-1/2}\sqrt{\beta_1\beta_2\beta_3}\Big(\frac{p}{\beta_2} -\frac{q}{\beta_3}\Big); \ell=\Big(\frac1\beta_1+\frac1\beta_2+\frac1\beta_3\Big)^2\beta_1\beta_2\beta_3.
\end{align} 
The Jacobian for this transformation is
\begin{equation}
	J^{-1} = \Big|\det{\frac{\partial \beta_i}{\partial s_j}}\Big|= \Big(\frac1\beta_1+\frac1\beta_2+\frac1\beta_3\Big) \frac{\beta_1\beta_2\beta_3}{\ell^2} = \frac{\sqrt{\beta_1\beta_2\beta_3}}{\ell^{3/2}}.
\end{equation}
We had done this substitution combining two transformations of variables. The first one so we could simplify the $k$ integral, second one so we could have the result in an easily identifiable standard integral.
\begin{align}
	&\mathcal{K}_{\mu\nu\rho}(p,q,r)\nonumber\\
	=&\int_{0}^\infty\mathrm{d}\beta_1\int_{0}^\infty\mathrm{d}\beta_2\int_{0}^\infty\mathrm{d}\beta_3 \int_{k'} \frac{\ell^{3/2}}{\sqrt{\beta_1\beta_2\beta_3}}\exp\{-\ell k'^2 -\beta_1(p+q)^2-\beta_2 q^2 -\beta_3 p^2\}\nonumber\\
	&\qquad(k'-xp+yq)_\mu(k'-xp+yq)_\nu(k'+(1-x)p+yq)_\rho,
\end{align}
where 
\begin{equation*}
	x\equiv \sqrt{\frac{\beta_1\beta_2\beta_3}{\ell}}\frac{1}{\beta_2};\quad y\equiv \sqrt{\frac{\beta_1\beta_2\beta_3}{\ell}}\frac{1}{\beta_3};\quad 1-x-y \equiv \sqrt{\frac{\beta_1\beta_2\beta_3}{\ell}}\frac{1}{\beta_1}.
\end{equation*}
Since this  integral couples to fields $a^\mu(-p)a^\nu(-q)a^\rho(-r)\dpqr$ which are transverse, and since $p^\mu+q^\mu+r^\mu=0$,
\begin{align*}
	&\mathcal{K}_{\mu\nu\rho}(p,q,r)\\
	=&\int_{0}^\infty\mathrm{d}\beta_1\int_{0}^\infty\mathrm{d}\beta_2\int_{0}^\infty\mathrm{d}\beta_3 \int_k \frac{\ell^{3/2}}{\sqrt{\beta_1\beta_2\beta_3}}\exp\{-\ell k'^2 -\beta_1r^2-\beta_2 q^2 -\beta_3 p^2\}\\
	&\qquad(k'+yq)_\mu(k'+xr)_\nu(k'+(1-x-y)p)_\rho.
\end{align*} 
The $k'$ integral only keeps the terms even in $k'$, and $\int_k f(k^2)k_\mu k_\nu=\dmn/D\int_k k^2f(k^2)$. In performing the $k'$ integral, note that
\begin{equation}
\ell^{3/2}\int\frac{\der^D k'}{(2\pi)^D} k'^n\exp\{-\ell k'^2\}=\frac{1}{2(2\pi)^D} \ell^{\frac32-\frac{D+n}2} \Gamma\Big(\frac{D+n}{2}\Big)
\end{equation}
\paragraph{Term 1}
\begin{align*}
	& q_\mu r_\nu p_\rho \int_{0}^\infty\mathrm{d}\beta_1\int_{0}^\infty\mathrm{d}\beta_2\int_{0}^\infty\mathrm{d}\beta_3 \int_k' \frac{\ell^{3/2}}{\sqrt{\beta_1\beta_2\beta_3}}xy(1-x-y)\exp\{-\ell k'^2 -\beta_1r^2-\beta_2 q^2 -\beta_3 p^2\}\\
	=&\frac{\Gamma(3/2)}{2(2\pi)^{3}}q_\mu r_\nu p_\rho\int_{0}^\infty\mathrm{d}\beta_1\int_{0}^\infty\mathrm{d}\beta_2\int_{0}^\infty\mathrm{d}\beta_3 \frac{1}{\sqrt{\beta_1\beta_2\beta_3}}xy(1-x-y)\exp\{ -\beta_1r^2-\beta_2 q^2 -\beta_3 p^2\}\\
	=&\frac{\Gamma(3/2)}{2(2\pi)^{3}}q_\mu r_\nu p_\rho\int_{0}^\infty\mathrm{d}\beta_1\int_{0}^\infty\mathrm{d}\beta_2\int_{0}^\infty\mathrm{d}\beta_3 \frac{1}{\ell^{3/2}}\exp\{ -\beta_1r^2-\beta_2 q^2 -\beta_3 p^2\}\\
	=&\frac{\Gamma(3/2)}{2(2\pi)^{3}}q_\mu r_\nu p_\rho\int_{0}^\infty\mathrm{d}\beta_1\int_{0}^\infty\mathrm{d}\beta_2\int_{0}^\infty\mathrm{d}\beta_3 \exp\{ -\beta_1r^2-\beta_2 q^2 -\beta_3 p^2\}(\beta_1\beta_2\beta_3)^{-3/2}\bigg(\frac1\beta_1+\frac1\beta_2+\frac1\beta_3\bigg)^{-3}
\end{align*}
\paragraph{Term 2}
\begin{align*}
	&\frac{1}{D}\frac{\Gamma(5/2)}{2(2\pi)^{3}}\int_{0}^\infty\mathrm{d}\beta_1\int_{0}^\infty\mathrm{d}\beta_2\int_{0}^\infty\mathrm{d}\beta_3 \int_k \frac{\ell^{3/2}}{\sqrt{\beta_1\beta_2\beta_3}}k^2\exp\{-\ell k^2 -\beta_1r^2-\beta_2 q^2 -\beta_3 p^2\}\\
	&\qquad((1-x-y)\dmn p_\rho+ yq_\mu \delta_{\nu\rho} +xr_\nu \delta_{\mu\rho})\\
	=&\frac{1}{D}\frac{\Gamma(5/2)}{2(2\pi)^{3}}\int_{0}^\infty\mathrm{d}\beta_1\int_{0}^\infty\mathrm{d}\beta_2\int_{0}^\infty\mathrm{d}\beta_3 \frac{\ell^{-1}}{\sqrt{\beta_1\beta_2\beta_3}}\exp\{ -\beta_1r^2-\beta_2 q^2 -\beta_3 p^2\}\\
	&\qquad((1-x-y)\dmn p_\rho+ yq_\mu \delta_{\nu\rho} +xr_\nu \delta_{\mu\rho})\\
	=&\frac{1}{D}\frac{\Gamma(5/2)}{2(2\pi)^{3}}\int_{0}^\infty\mathrm{d}\beta_1\int_{0}^\infty\mathrm{d}\beta_2\int_{0}^\infty\mathrm{d}\beta_3 \ell^{-3/2}\exp\{ -\beta_1r^2-\beta_2 q^2 -\beta_3 p^2\}\\
	&\qquad(\frac{1}{\beta_1}\dmn p_\rho+ \frac1{\beta_3}q_\mu \delta_{\nu\rho} +\frac1{\beta_2}r_\nu \delta_{\mu\rho})\\
	=&\frac{1}{D}\frac{\Gamma(5/2)}{2(2\pi)^{3}}\int_{0}^\infty\mathrm{d}\beta_1\int_{0}^\infty\mathrm{d}\beta_2\int_{0}^\infty\mathrm{d}\beta_3 \bigg(\frac1\beta_1+\frac1\beta_2+\frac1\beta_3\bigg)^{-3}(\beta_1\beta_2\beta_3)^{-3/2}\\
	&\qquad\exp\{ -\beta_1r^2-\beta_2 q^2 -\beta_3 p^2\} \Big(\frac{1}{\beta_1}\dmn p_\rho+ \frac1{\beta_3}q_\mu \delta_{\nu\rho} +\frac1{\beta_2}r_\nu \delta_{\mu\rho}\Big)
\end{align*}
We use the following integral.
\begin{equation}
	2^{-2a}\int_{0}^\infty \der X \frac{X^{a-1}}{\Gamma(a)}e^{-XT/4}=T^{-a}.
\end{equation} 
Then
\begin{align*}
&\text{Term 1}\\	
=&\frac{\Gamma(3/2)}{2^7\Gamma(3)(2\pi)^{3}}q_\mu r_\nu p_\rho\int_{0}^\infty\mathrm{d}\beta_1\int_{0}^\infty\mathrm{d}\beta_2\int_{0}^\infty\mathrm{d}\beta_3 \int_{0}^\infty \mathrm{d}XX^2(\beta_1\beta_2\beta_3)^{-3/2}\exp\bigg\{ -\beta_1r^2-\beta_2 q^2\\
& \qquad-\beta_3 p^2-\frac{X}{4}\bigg(\frac1\beta_1+\frac1\beta_2+\frac1\beta_3\bigg)\bigg\}
\end{align*}
\begin{align*}
&\text{Term 2}
\\
	=&\frac{1}{D}\frac{\Gamma(5/2)}{2^7\Gamma(3)(2\pi)^{3}}\int_{0}^\infty\mathrm{d}\beta_1\int_{0}^\infty\mathrm{d}\beta_2\int_{0}^\infty\mathrm{d}\beta_3\int_{0}^\infty \der X \frac{X^{2}}{\Gamma(3)} (\beta_1\beta_2\beta_3)^{-D/2}\exp\bigg\{ -\beta_1r^2-\beta_2 q^2 \nonumber\\
	&\qquad-\beta_3 p^2 -\frac X4\bigg(\frac1\beta_1+\frac1\beta_2+\frac1\beta_3\bigg)\bigg\} \Big(\frac{1}{\beta_1}\dmn p_\rho+ \frac1{\beta_3}q_\mu \delta_{\nu\rho} +\frac1{\beta_2}r_\nu \delta_{\mu\rho}\Big).
\end{align*}
These $\beta$ integrals are of the form
\begin{equation}
	\int_{0}^{\infty}\der \beta \beta^{-\nu-1}\exp\bigg\{-k^2\beta-\frac{X}{4\beta} \bigg\} = 2^{1+\nu}(k/\sqrt X)^{\nu} K_\nu(k\sqrt X).
\end{equation}
In this integral, $X$ acts as the UV cutoff for $\beta$, which is of the dimension $-2$, and $k^2$ acts as the IR cutoff. At this stage, we identify the lower bound of the $X$ integral with the inverse square of the UV cutoff $1/\Lambda^2=z^2$.
Thus we have, (apologies for using the same symbol $\nu$ as the quantity $\Delta-D/2=1/2$ and as a spacetime index for the momenta),
\begin{align*}
&\text{Term 1}\\
=&\frac{2^{3\nu}\Gamma(3/2)}{2^5(2\pi)^{3}}q_\mu r_\nu p_\rho \int_{z^2}^\infty \mathrm{d}X{X^2} \bigg(\frac p{\sqrt X}\bigg)^{\nu} K_\nu(p\sqrt X)\bigg(\frac q{\sqrt X}\bigg)^{\nu} K_\nu(q\sqrt X) \bigg(\frac{r}{\sqrt X}\bigg)^{\nu} K_\nu(r\sqrt X).
\end{align*}
\begin{align*}
	&\text{Term 2}\\
	=& \frac{1}{D}\frac{2^{3\nu}\Gamma(5/2)}{2^4(2\pi)^{3}}\int_{z^2}^\infty \der X \frac{X^{2}}{2}\Big[\dmn p_\rho\Big(\frac{r}{\sqrt X}\Big)^{1+\nu}\Big(\frac{p}{\sqrt X}\Big)^{\nu}\Big(\frac{q}{\sqrt X}\Big)^{\nu}K_{1+\nu}(r\sqrt X)K_{\nu}(p\sqrt X)K_{\nu}(q\sqrt X) \nonumber\\
	&+q_\mu \delta_{\nu\rho} \Big(\frac{r}{\sqrt X}\Big)^{\nu}\Big(\frac{p}{\sqrt X}\Big)^{1+\nu}\Big(\frac{q}{\sqrt X}\Big)^{\nu}K_{\nu}(r\sqrt X)K_{1+\nu}(p\sqrt X)K_{\nu}(q\sqrt X)\nonumber\\
	&+r_\nu \delta_{\mu\rho} \Big(\frac{r}{\sqrt X}\Big)^{\nu}\Big(\frac{p}{\sqrt X}\Big)^{\nu}\Big(\frac{q}{\sqrt X}\Big)^{1+\nu}K_{\nu}(r\sqrt X)K_{\nu}(p\sqrt X)K_{1+\nu}(q\sqrt X)\Big].
	\label{term2-integral}
\end{align*}
In the above, when the momenta, $p,q,r$ appear without their index, it refers to their magnitude.

The derivative of this with respect to $z$ appears in our cubic term.
\begin{align}
\ddz \mathcal{K}&_{\mu\nu\rho}(p,q,r;z)\nonumber\\
=&-\frac{2^{3\nu}\Gamma(3/2)}{2^4(2\pi)^{3}}z^5q_\mu r_\nu p_\rho\Big(\frac{r}{z}\Big)^{\nu}\Big(\frac{p}{z}\Big)^{\nu}\Big(\frac{q}{z}\Big)^{\nu}K_{\nu}(rz)K_{\nu}(pz)K_{\nu}(qz) \nonumber\\
&-\frac{1}{D}\frac{2^{3\nu}\Gamma(5/2)}{2^3(2\pi)^{3}} z^{5}\Big[\dmn p_\rho\Big(\frac{r}{z}\Big)^{1+\nu}\Big(\frac{p}{z}\Big)^{\nu}\Big(\frac{q}{z}\Big)^{\nu}K_{1+\nu}(rz)K_{\nu}(pz)K_{\nu}(qz) \nonumber\\
&+q_\mu \delta_{\nu\rho} \Big(\frac{r}{z}\Big)^{\nu}\Big(\frac{p}{z}\Big)^{1+\nu}\Big(\frac{q}{z}\Big)^{\nu}K_{\nu}(rz)K_{1+\nu}(pz)K_{\nu}(qz)\nonumber\\
&+r_\nu \delta_{\mu\rho} \Big(\frac{r}{z}\Big)^{\nu}\Big(\frac{p}{z}\Big)^{\nu}\Big(\frac{q}{z}\Big)^{1+\nu}K_{\nu}(rz)K_{\nu}(pz)K_{1+\nu}(qz)\Big].
\end{align}
We make use of the following property of modified Bessel functions\footnote{See the entry at Wolfram Functions site. https://web.archive.org/web/20081013054654/https://functions.wolfram.com/Bessel-TypeFunctions/BesselK/introductions/Bessels/05/} to reduce the parameter of the Bessel functions to $\nu$. 
\begin{equation}
	K_{1+\nu}(x)=K_{\nu-1}(x)+\frac{2\nu}{x}K_\nu(x).
\end{equation}
For our purposes, $\nu=1/2$, ($\because K_{1/2}(x)=K_{-1/2}(x)$ ),
\begin{equation}
	K_{3/2}(pz)=(1+1/pz)K_{1/2}(pz).
\end{equation}
Thus
\begin{align}
		\ddz \mathcal{K}&_{\mu\nu\rho}(p,q,r;z)\nonumber\\
		=&-\frac{1}{16(2\pi)^{5/2}}z^5q_\mu r_\nu p_\rho\Big(\frac{r}{z}\Big)^{\nu}\Big(\frac{p}{z}\Big)^{\nu}\Big(\frac{q}{z}\Big)^{\nu}K_{\nu}(rz)K_{\nu}(pz)K_{\nu}(qz) \nonumber\\
		&-\frac{1}{16(2\pi)^{5/2}} z^{5}\Big[\dmn p_\rho\Big(\frac{r}{z}\Big)^{1+\nu}\Big(\frac{p}{z}\Big)^{\nu}\Big(\frac{q}{z}\Big)^{\nu}\Big(1+\frac{2\nu}{rz}\Big)+q_\mu \delta_{\nu\rho} \Big(\frac{r}{z}\Big)^{\nu}\Big(\frac{p}{z}\Big)^{1+\nu}\Big(\frac{q}{z}\Big)^{\nu}\Big(1+\frac{2\nu}{pz}\Big)\nonumber\\
		&-p_\nu \delta_{\mu\rho} \Big(\frac{r}{z}\Big)^{\nu}\Big(\frac{p}{z}\Big)^{\nu}\Big(\frac{q}{z}\Big)^{1+\nu}\Big(1+\frac{2\nu}{qz}\Big)\Big]K_{\nu}(rz)K_{\nu}(pz)K_{\nu}(qz)\nonumber\\
		=&-\frac{1}{16(2\pi)^{5/2}} z^{3}\Big[q_\mu r_\nu p_\rho z^2+\dmn p_\rho(1+{rz})
		+q_\mu \delta_{\nu\rho}(1+{pz})+r_\nu \delta_{\mu\rho} (1+{qz})\Big]\nonumber\\
		&\Big(\frac{r}{z}\Big)^{\nu}\Big(\frac{p}{z}\Big)^{\nu}\Big(\frac{q}{z}\Big)^{\nu}K_{\nu}(rz)K_{\nu}(pz)K_{\nu}(qz).
		\label{localResultOfIntegral}
\end{align}

In \ref{mapping-cubic}, we see that the factor $\Big(\frac{r}{z}\Big)^{\nu}\Big(\frac{p}{z}\Big)^{\nu}\Big(\frac{q}{z}\Big)^{\nu}K_{\nu}(rz)K_{\nu}(pz)K_{\nu}(qz)$ is cancelled. 

\section{Boundary 3-pt function from the bulk action}
In this appendix we verify that the 3-point correlator calculated by the bulk AdS action in \eqref{gaugeinvariantaction} obeys the conformal ward identities by matching it with \cite{Bzowski:2013sza}.

In section \ref{GaugeInvariance}, we showed the action \eqref{gaugeinvariantaction} is equivalent to the action in \eqref{cubicTerm} on-shell,
\begin{equation}
	S_3 
	=-\eps_{ijk}\int\mathrm dz \int_p\int_q\int_r \dpqr [2g'z^2q_\mu r_\nu p_\rho\amup(p) a^{\nu}_j(q) a^{\rho}_k(r)
	+(g+6g'{|r|z})
	\amup(p) a_{\mu}^j(q)\times p_\rho a^{\rho}_k(r)].
	\label{actionreduced}
\end{equation}

The solution to the equations of motion to $O(g^0,g'^0)$, is given in \eqref{bulksolution}.
\begin{equation}
	\bar{a}_{\mu}^i(p,z)=A_\mu^i(p)\sqrt {|p|z} K_{\frac12}(pz),
\end{equation}
where $A_\mu^i(p)$ is the boundary source. Substituting this into \eqref{actionreduced}, we have
\begin{align}
	S_{\textrm{cl.}3}=&-\eps_{ijk} \int_p\int_q\int_r \dpqr A^\mu_i(p) A^{\nu}_j(q) A^{\rho}_k(r)\sqrt{|p||q||r|}\nonumber\\ &\bigg[2g'q_\mu r_\nu p_\rho  \int\mathrm dz z^{7/2} K_{\frac12}(pz)K_{\frac12}(qz)K_{\frac12}(rz)+ p_\rho \dmn\Big(g \int\mathrm dz z^{3/2} K_{\frac12}(pz)K_{\frac12}(qz)K_{\frac12}(rz)\nonumber\\
	&+6g'{|r|} \int\mathrm dz z^{5/2} K_{\frac12}(pz)K_{\frac12}(qz)K_{\frac12}(rz)\Big)\bigg].
\end{align}
The integrals of the bessel functions can be computed to get
\begin{align}
	S_{\textrm{cl.}3}=&-\eps_{ijk}\Big(\frac{\pi}{ 2}\Big)^{3/2} \int_p\int_q\int_r \dpqr A^\mu_i(p) A^{\nu}_j(q) A^{\rho}_k(r)\bigg[4g'q_\mu r_\nu p_\rho  \frac{1}{(|p|+|q|+|r|)^3}\nonumber\\ 
	&+ p_\rho \dmn\Big( \frac g{|p|+|q|+|r|} +6g'\frac{|r|}{(|p|+|q|+|r|)^2} \Big)\bigg].
\end{align}
This matches with the result for 3 point functions obeying conformal ward identities in section 9.5 of \cite{Bzowski:2013sza}.

%%%%%%%%%%%%%%%%%%%%%%%%%%%%%%%%%%%%%%%%%%%%%%%%%%%%%%%%%%%%%%%%%%%%%%%%%%%%%%%
\pagebreak
\bibliography{holographicRG.bib,AdS-CFT.bib,higherspins.bib,erg.bib}
\bibliographystyle{utphys}
\end{document}